\documentclass{edm_article}
\usepackage{graphicx}
\usepackage{textcomp}
\usepackage[table]{xcolor}
\usepackage{enumerate}% http://ctan.org/pkg/enumerate
\usepackage{subcaption}
\usepackage{enumitem}
\usepackage{amsmath}
\usepackage{booktabs}
\usepackage{array} 
\usepackage[protrusion=true,expansion=true]{microtype}
\usepackage{multirow}
\usepackage{enumitem}

\begin{document}

\title{Privacy-Preserving Distributed Link Predictions Among Peers in Online Classrooms Using Federated Learning}
\numberofauthors{6}
\author{
\alignauthor
\hspace{-1mm}
Anurata Prabha Hridi \\
       % \affaddr{\hspace{-2mm}Computer Science}\\
       \affaddr{\hspace{-2mm}NC State University}\\
       % \affaddr{\hspace{-2mm}North Carolina, USA}\\
       \email{\hspace{-4mm} aphridi@ncsu.edu}
% 2nd. author
% \hspace{-5mm}
\alignauthor
\hspace{-15mm} Muntasir Hoq\\ 
       % \affaddr{\hspace{-14mm}Computer Science}\\
       \affaddr{\hspace{-14mm}NC State University}\\
       % \affaddr{\hspace{-14mm}North Carolina, USA}\\
       \email{\hspace{-14mm}mhoq@ncsu.edu}
% 3rd. author
\alignauthor \hspace{-35mm}Zhikai Gao\\
       % \affaddr{\hspace{-35mm}Computer Science}\\
       \affaddr{\hspace{-35mm}NC State University}\\
       % \affaddr{\hspace{-35mm}North Carolina, USA}\\
       \email{\hspace{-35mm}zgao9@ncsu.edu}
% 4th Author
\alignauthor \hspace{-55mm} Collin Lynch\\
       % \affaddr{\hspace{-55mm}Computer Science}\\
       \affaddr{\hspace{-55mm}NC State University}\\
       % \affaddr{\hspace{-55mm}North Carolina, USA}\\
       \email{\hspace{-55mm} cflynch@ncsu.edu }
\and  % use '\and' if you need 'another row' of author names
% 5th. author
\alignauthor Rajeev Sahay\\
       % \affaddr{Electrical and Computer Engineering}
       \affaddr{UC San Diego}
       % \affaddr{California, USA}
       \email{r2sahay@ucsd.edu}
% 6th. author
\alignauthor Seyyedali Hosseinalipour\\
       % \affaddr{Electrical Engineering}
       \affaddr{University at Buffalo--SUNY}
       % \affaddr{New York, USA}
       \email{alipour@buffalo.edu}
% 7th. author
\alignauthor Bita Akram\\
       % \affaddr{Computer Science}\\
       \affaddr{NC State University}\\
       % \affaddr{North Carolina, USA}\\
       \email{bakram@ncsu.edu}
}
\maketitle
\begin{abstract}
Social interactions among classroom peers, represented as social learning networks (SLNs), play a crucial role in enhancing learning outcomes. While SLN analysis has recently garnered attention, most existing approaches rely on centralized training, where data is aggregated and processed on a local/cloud server with direct access to raw data. However, in real-world educational settings, such direct access across multiple classrooms is often restricted due to privacy concerns. Furthermore, training models on isolated classroom data prevents the identification of common interaction patterns that exist across multiple classrooms, thereby limiting model performance. To address these challenges, we propose one of the first frameworks that integrates Federated Learning (FL), a distributed and collaborative machine learning (ML) paradigm, with SLNs derived from students' interactions in multiple classrooms’ online forums to predict future link formations (i.e., interactions) among students. By leveraging FL, our approach enables collaborative model training across multiple classrooms while preserving data privacy, as it eliminates the need for raw data centralization. Recognizing that each classroom may exhibit unique student interaction dynamics, we further employ model personalization techniques to adapt the FL model to individual classroom characteristics. Our results demonstrate the effectiveness of our approach in capturing both shared and classroom-specific representations of student interactions in SLNs. Additionally, we utilize explainable AI (XAI) techniques to interpret model predictions, identifying key factors that influence link formation across different classrooms. These insights unveil the drivers of social learning interactions within a privacy-preserving, collaborative, and distributed ML framework—an aspect that has not been explored before.

\end{abstract}

 % Replace with your own 3-5 keywords

\section{Introduction}
Students' social interactions, a founding factor of learning according to the \textit{socio-constructivist theory of learning}~\cite{kanselaar2002constructivism}, hold special importance in an online/hybrid learning environment.
Unlike traditional settings, where peer-to-peer and peer-to-instructor interactions occur naturally, these environments often rely on limited and deliberately designed interaction channels~\cite{wut2021person}. 
This shift in interaction dynamics, shaped by the widespread adoption of modern learning platforms such as massive open online courses (MOOCs) and learning management systems (LMSs), not only alters the way students engage but also provides a structured setting for systematically examining these interactions~\cite{rubin2010effect}. Understanding how this shift affects the nature and quality of interactions is crucial, as these interactions play a direct role in shaping students' learning experiences and outcomes.  To this end, researchers have drawn parallels between student interactions in online learning environments and user connections in social networks, representing them through \textit{graph-based data structures} known as social learning networks (SLNs) \cite{brinton2018efficiency, gurjar2020leveraging}. Analyzing SLNs uncovers valuable insights into interaction patterns and their correlation with student learning, enabling the development of personalized instructional strategies that foster engagement, enhance learning outcomes, and promote academic success \cite{carolan2013social}.
% The prevalent use of modern learning platforms, including massive open online courses (MOOCs) and learning management systems (LMSs) have brought about an exceptional opportunity for a deeper investigation of students' interactive behavior and its effect on their learning experiences and outcomes \cite{rubin2010effect}. Similar to user interactions in a social network, students' interactions with their peers can be represented by graph-based data structures called social learning networks (SLNs) \cite{brinton2018efficiency, gurjar2020leveraging}. Analyzing SLNs sheds light on students' interaction patterns and their potential connection with students' learning. This can eventually offer personalized instructional support to promote learning, engagement, and success \cite{carolan2013social}. 

\vspace{-4mm}
\subsection{Background and Motivation}
Historically, artificial intelligence/machine learning (AI/ML) techniques have been utilized to analyze SLNs and understand the complexities of social interactions among students \cite{bashiri2024transformative, williams2021social}. 
Despite significant progress, the existing works in this domain share a common limitation: \textit{a reliance on centralized learning architectures, where ML models are trained on local/cloud servers with direct access to all training data~\cite{choudhary2023social}.} Such direct data access typically occurs under two scenarios: (i) data from multiple classrooms are aggregated into a centralized storage system or (ii) model training is performed using data from a single classroom. Nevertheless, while the former approach enhances model performance through access to diverse and high-quality data, it is often restricted by privacy regulations -- such as family educational rights and privacy act (FERPA)~\cite{FERPA} -- that prohibit the transfer of student data across networks due to the risk of data exposure. Also, the latter approach, although privacy-compliant, generally leads to suboptimal model performance due to the limited diversity and volume of available training data. 
% Prior work overwhelmingly relies on centralized methods, where ML models are trained by directly accessing and aggregating data available across multiple sources \cite{choudhary2023social}. However, due to privacy concerns \cite{kairouz2021advances}% and resource constraints \cite{carbajo2024adaptive}
% , availing such raw data for training purposes cannot be practiced in the education domain. Therefore, centralized models are often trained on a subset of available data, typically from a single classroom or institution \cite{jalil2019machines}. However, this approach banks on isolated training and fails to capture key patterns existent in the collective data. %, and second, a lack of fairness across classrooms due to bias against underrepresented students, whose limited data representation hinders effective model training. 

Recently, federated learning (FL) has emerged as a leading distributed ML technique \cite{mcmahan2017communication, kairouz2021advances}, offering a viable solution to the limitation discussed above. In particular, FL enables collaborative model training across distributed data-collecting entities (referred to as clients) through an iterative process that involves two primary steps. \textit{(Step 1) Local Training:} each client independently trains a local model on its dataset and shares only the model parameters (rather than raw data) with a central server. \textit{(Step 2) Global Aggregation and Broadcast:} the server aggregates the received local model parameters from clients --- typically through weighted averaging --- to construct a global model, which is subsequently broadcast back to the clients, allowing them to synchronize their local models and commence the next round of local training.
By eliminating the need to transfer raw data across networks, FL inherently preserves privacy while facilitating distributed ML across multiple clients. This privacy-preserving property allows FL to enhance both the performance and fairness of model training compared to centralized methods, which rely on isolated data from individual clients \cite{asad2021federated, peng2022centralized, tungar2023evaluation, jiang2022privacy, khelghatdoust2022socially}. Given these advantages, FL has recently gained significant attention in various domains, including educational applications \cite{chu2024multi, chu2022mitigating, sengupta2024fedel, bhattacharya2023towards, ebrahimi2025transition} and social networks \cite{yang2018predicting, baek2023personalized, wu2021fedgnn, ammad2019federated, liu2022federated, khelghatdoust2022socially}. Despite these advances, \textit{the application of FL for harnessing data from multiple classrooms to study SLNs through the lens of predicting student interactions (referred to as link prediction) remains largely unexplored}. Addressing this gap and investigating the performance of FL in this context serves as the primary motivation for this paper.

\vspace{-1mm}
\subsection{Overview and Summary of Contributions}
Our key contributions can be summarized as follows:
\vspace{-3mm}

\begin{itemize}[leftmargin=4mm]
\setlength\itemsep{-.1em}
    \item To our knowledge, this is the first study that explores the potential of FL, along with its advanced variations, to predict the \textit{occurrence} (or absence) of an \textit{interaction} between students (represented as an \textit{edge} connecting two nodes in an SLN) based on their interaction patterns. Consequently, this work establishes the framework for privacy-preserving distributed link prediction in SLNs.
    
    \item 
   To conduct our analysis, we use forum interaction data from five distinct classrooms focused on science and humanities, from which we extract a set of graph-theoretic features. This process reveals a unique challenge in studying FL performance for SLN analysis: \textit{while student interaction behaviors may share commonalities across classrooms, each class --- due to its unique subject matter and structure --- often exhibits distinct interaction patterns}. Nevertheless, vanilla FL methods, which train a single/universal global model across all classrooms, fail to capture these distinct interaction patterns.
    
    \item To address this challenge, we investigate a set of \textit{model personalization strategies} to adapt the global FL model to the data of each classroom. Our findings demonstrate the superior performance of fine-tuned/personalized FL models over both centralized and vanilla FL approaches, achieving notable improvements in both \textit{classification accuracy} and \textit{prediction fairness} metrics. These results highlight the effectiveness of personalization in accounting for classroom-specific differences and pave the way for future work on further personalization of FL models based on student demographics, promoting more inclusive and representative educational environments.
    
\item We then investigate an understudied aspect: \textit{whether the importance of graph-theoretic features varies between science and humanities classrooms due to their distinct interaction patterns.} To this end, we incorporate explainable AI (XAI) techniques --- specifically, SHapley Additive exPlanations (SHAP) --- into the analysis of fine-tuned/personalized FL models. This examination provides nuanced insights into the varying influence of graph-theoretic features on student interaction predictions across different classroom types, enabling the development of tailored recommendations for educators and institutions to foster meaningful student connections based on course subject and classroom context.
\end{itemize}

\vspace{-3mm}
\section{Related Work}
% SLNs provide valuable insights into student collaboration and engagement \cite{xu2018many}, and machine learning (ML) techniques have been widely used to predict connections within these networks, promoting personalized learning and enhancing peer knowledge exchange. While federated learning (FL) has seen successful applications in various graph-based tasks \cite{baek2023personalized}, its potential for link prediction in social networks within educational contexts remains largely unexplored. 
In the following, we first provide an overview of social learning and the use of AI/ML techniques in this domain, and then we delve deeper into the link prediction problem in SLNs. 

% \vspace{-1mm}
\subsection{Social Learning \& the Impact of AI/ML}
Social learning in traditional educational settings emphasizes peer interactions, which has been shown to result in enhanced cognitive skills \cite{bandura1977social}. In this process, students acquire knowledge and develop competencies through social interactions \cite{kanselaar2002constructivism}. In recent years, the rise of online learning platforms such as Coursera and edX has significantly expanded the reach of social learning, leading to the formation of SLNs. 
These platforms support virtual communities and discussion forums, enabling learners to connect, collaborate, and share resources across diverse geographical locations.
Also, the growth of online education, particularly accelerated by the COVID-19 pandemic, has led to the widespread adoption of digital learning platforms, with nearly 84\% of undergraduate students in the US engaging in at least one online course \cite{hollister2022engagement}. These online/digital platforms generate extensive data through student activity logs and peer interactions, which can be modeled as SLNs \cite{xu2018many}, where \textit{nodes} represent students and \textit{edges} denote interactions. 
With the advancement of AI/ML techniques, analyzing student activity logs from these platforms has been shown to uncover meaningful interaction behaviors, creating opportunities to enhance the learning environment \cite{amershi2009combining}. 
By capturing diverse patterns within data, ML models can address learners' individual needs, making SLNs a powerful tool for increasing engagement and offering personalized learning recommendations \cite{gurjar2020leveraging}.  
% Therefore, the intersection of social learning with emerging AI/ML technologies has unlocked new possibilities for studying and improving digital education platforms. These technologies facilitate the personalization of social learning experiences by analyzing interaction patterns and adapting content to meet individual learner needs.
% By capturing diverse patterns within data, ML models can address learners' individual needs, making SLNs a powerful tool for increasing engagement and offering personalized recommendations \cite{hasan2011survey, gurjar2020leveraging}.  

% \vspace{-1mm}
\subsection{Link Prediction in SLN \& Potential of FL}
Link prediction has been studied in the context of both SLNs and online social networks (OSNs) --- which bear a strong resemblance to SLNs --- for decades \cite{hasan2011survey, liben2003link, brinton2018efficiency, mezghani2018online}, primarily leveraging neighborhood and path-based features derived from graph topologies \cite{muniz2018combining, jie2019social, backstrom2011supervised}. In this domain, various ML methods have been used to improve the accuracy of link predictions \cite{aghababaei2019interpolative}. Nevertheless, the existing approaches in this domain rely on \textit{centralized} model training procedures, which suffer from either limited data exposure \cite{gharahighehi2022addressing} when trained on isolated data of an individual OSN/SLN or face privacy concerns when the model is trained on the pool of data collected from multiple OSNs/SLNs \cite{liu2022federated}. This hinders the effectiveness of models derived from OSNs/SLNs in promoting collaboration and personalized learning experiences. 

FL offers a solution by enabling distributed model training across multiple clients while keeping data localized, thereby ensuring privacy and addressing data limitations \cite{mcmahan2017communication, liu2022distributed}. Subsequently, by aggregating knowledge from various classrooms, each considered to be a client in the FL setting, FL enhances the generalizability and accuracy of the models trained on multiple classrooms' data. While recent studies have examined the applications of FL in educational settings \cite{chu2022mitigating, sengupta2024fedel, bhattacharya2023towards, wu2021federated, hridi2024revolutionizing}, its potential for link prediction in SLNs remains unexplored --- a gap that this work aims to fill.

% \vspace{-1mm}
\section{Methodology}
In this section, we first describe the SLN structure and the datasets involved in this study~(Sec.~\ref{sec:3-a}). We then elaborate on the centralized learning~(Sec.~\ref{sec:3-b}) and the FL architecture~(Sec.~\ref{sec:3-c}). Afterward, we describe three personalized FL (pFL) methods that entail tailoring the FL model to individual classrooms (Sec.~\ref{sec:3-d}). Finally, we delve into the fairness measures and XAI techniques~(Secs.~\ref{sec:3-e} and \ref{sec:3-f}). 
%In this section, the SLN structure, datasets, and model architectures involved in this study are described. The centralized local learning methods and FL models are trained and evaluated with the data from SLNs for link prediction. %We used Accuracy, Loss, AUC, and Equalized Odds measures to evaluate model performance and fairness. 

\vspace{-1.5mm}
\subsection{Social Learning Network (SLN) Structure}\label{sec:3-a}
% Before diving into our data and method to predict links, 

% \vspace{-4mm}
\begin{figure}[t]
    \centering
    \begin{subfigure}[b]{0.24\textwidth}
        \centering
    \includegraphics[width=1.032\linewidth, alt={SLNs illustrating students' connections formed at time $t-1$, where, $S_1$, $S_2$, etc., represent individual students. Out of 10 students, four stay unconnected by the end of $t-1$.}]{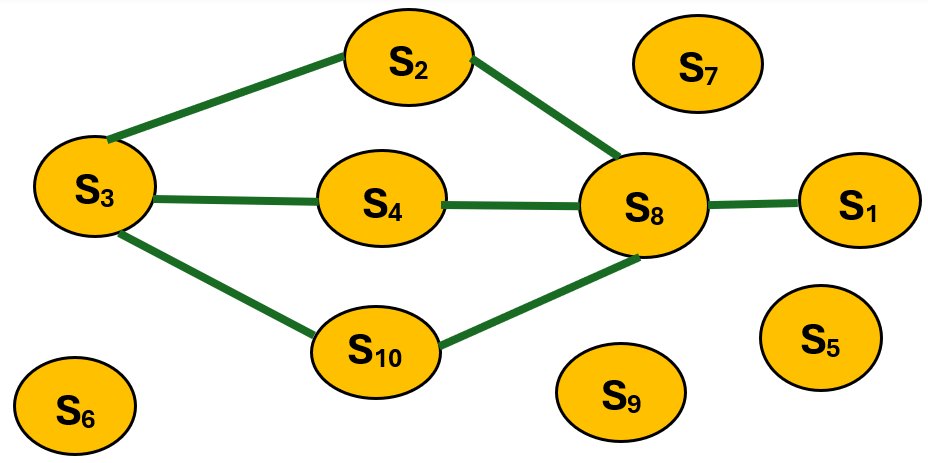}
    \caption{At time $t-1$}
    \end{subfigure}
    % \hfill
    \hspace{0.095cm}
    \begin{subfigure}[b]{0.22\textwidth}
        \centering
        \includegraphics[width=1.03\linewidth, alt={SLNs illustrating students' connections formed at time $t$, where, $S_1$, $S_2$, etc., represent individual students. Besides existing connections, new links were formed by the end of $t$ that can be attributed to course structure and student interactions.}]{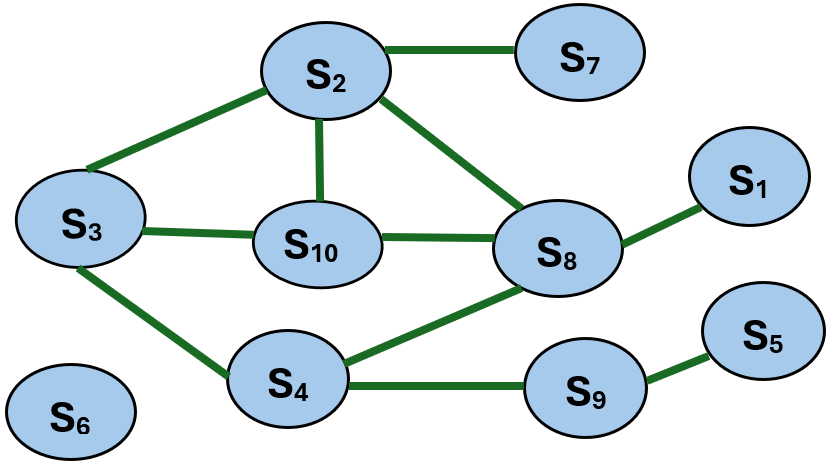}
        \caption{At time $t$}
    \end{subfigure}
    \caption{SLNs illustrating students' connections formed at two subsequent time instants, where, $S_1$, $S_2$, etc., represent individual students. (a) Out of 10 students, four stay unconnected by the end of $t-1$, and (b) besides existing connections, new links were formed by the end of $t$ that can be attributed to course structure and student interactions.} %In temporal networks, paths are not static \cite{holme2012temporal} so connections can change from time to time, as shown here.}
    \label{fig:sln}
\end{figure}
% \vspace{-0.2cm}
We begin by describing how SLNs are constructed based on student pairs and their shared contributions to various online activities, including discussions, resource sharing, and collaborative projects. 
Figure \ref{fig:sln} shows the temporal progression of a sample SLN formed from student interaction data, where nodes represent students and edges represent connections or interactions between them in the SLN. Let $t=0$ denote the initial time instant corresponding to the start of a course, and let $\mathcal{S} = \{S_1, S_2, \cdots\}$ represent the set of enrolled students. At any given time instant $t$, a link $(u, v)$ is established between two students $S_u$ and $S_v$, where $S_u, S_v \in \mathcal{S}$, if they have interacted prior to time $t$. Such interactions include initiating a common thread or adding follow-up posts in the same thread \cite{sahay2023predicting} and are depicted as bidirectional edges in the SLN to reflect the reciprocal nature of the communication.
To model the evolving SLN, we construct a time-varying adjacency matrix $\mathbf{A}^{(t)} = [a^{(t)}_{u,v}]_{S_u,S_v \in \mathcal{S}}$, where each entry $a^{(t)}_{u,v} \in \{0,1\}$ is a binary indicator denoting the presence ($a^{(t)}_{u,v} = 1$) or absence ($a^{(t)}_{u,v} = 0$) of an edge between students $S_u$ and $S_v$ at time $t$. Accordingly, the SLN at time instant $t$ can be represented by the time-dependent graph structure $\mathcal{G}^{(t)} = (\mathcal{S}, \mathcal{E}^{(t)})$, where $\mathcal{S}$ denotes the set of nodes (students) and $\mathcal{E}^{(t)}$ is the set of edges. A pair $(u, v)$ belongs to $\mathcal{E}^{(t)}$ (i.e., $(u, v) \in \mathcal{E}^{(t)}$) if and only if $a^{(t)}_{u,v} = 1$.

\vspace{-1mm}
\subsubsection{Datasets}
We consider four MOOC datasets adopted by Sahay et al. \cite{sahay2023predicting} and one real-world classroom data of Fall 2021 from a North American university. These datasets encompass a broad spectrum of subjects: \textit{Algorithms: Design and Analysis, Part 1 (algo), English Composition I (comp), Shakespeare in Community (shake), Machine Learning (ml)} and \textit{Software Development Fundamentals (csc)}.
When comparing the datasets based on the number of considered student pairs (i.e, student pairs whose interactions were recorded), \textit{`ml'} is the largest, followed by \textit{`algo'}, \textit{`shake'}, \textit{`comp'}, and \textit{`csc'}, which is the smallest (see Table~\ref{tab:details}). The relatively small number of links compared to the student pairs indicates sparsity of
the SLNs. These datasets cover diverse disciplines, comprising three STEM subjects and two humanities subjects, with no overlap (i.e., no shared enrolled students) in the collected data. 
% \anurata{}{OR: These datasets cover diverse disciplines —-- three from STEM and two from the humanities. Their data sources being different, each dataset contains different students, with no possible overlap in the collected data.}
In the context of constructing SLN graphs, each dataset includes links representing interactions formed among students only \cite{brown2015good} throughout the course duration. It is important to note that, in adherence to standard data anonymization practices, the datasets do not contain user-identifying information or student demographic details. 
\vspace{-0.4cm}

\begin{table}[h]
\centering
\caption{Course details and dataset characteristics.\vspace{-0.3cm}}
{\small
\begin{tabular}{|>{\centering\arraybackslash}m{1.85cm}||c|>{\centering\arraybackslash}m{0.5cm}|>{\centering\arraybackslash}m{0.5cm}|>{\centering\arraybackslash}m{0.8cm}|>{\centering\arraybackslash}m{1.6cm}|}
% \begin{tabular}{|c{1.2cm}|c|c{.3cm}|c|c|}
% \begin{tabular}{|c||c|c|c|c|}
\hline
\textbf{Course} & \textbf{Class} & \multirow{2}{*}{\hspace{-1.1mm}\textbf{Nodes}} &\multirow{2}{*}{\hspace{-0.5mm}\textbf{Links}} &\textbf{Student} & \textbf{Course}\\ 
\textbf{Title}&\textbf{-room}&&&\textbf{Pairs} & \textbf{Type}\\ 
 \hline  \hline
 %\rowcolor{green!20} % First row color
\cellcolor{green!20} Algorithms & \multirow{3}{*}{algo} & \multirow{3}{*}{2080} &\multirow{3}{*}{5416} & \multirow{3}{*}{36390} & STEM\\ 
\cellcolor{green!20} Design &&&&& \&\\ 
\cellcolor{green!20} Analysis I &&&&&MOOC\\ 
\hline
%\rowcolor{red!20} % First row color 
\cellcolor{red!20} English & \multirow{2}{*}{comp} & \multirow{2}{*}{1707}& \multirow{2}{*}{2731} & \multirow{2}{*}{19049} & Non-STEM\\ 
\cellcolor{red!20} Composition I &&&&&\& MOOC\\ 
\hline
%\rowcolor{blue!20} % First row color
\cellcolor{blue!20} Shakespeare& \multirow{2}{*}{shake} & \multirow{2}{*}{1296}&\multirow{2}{*}{3753} & \multirow{2}{*}{26008} & Non-STEM\\
\cellcolor{blue!20} in Community& &&&&\& MOOC\\ 
\hline
%\rowcolor{orange!20} % First row color
\cellcolor{orange!20} Machine& \multirow{2}{*}{ml} & \multirow{2}{*}{6446} &\multirow{2}{*}{\hspace{-0.7mm}24481} & \multirow{2}{*}{76526} & STEM\\
\cellcolor{orange!20} Learning &&&&&\& MOOC\\ 
\hline
%\rowcolor{orange!20} % First row color
\cellcolor{gray!20} Software& \multirow{3}{*}{csc} & \multirow{3}{*}{366}&\multirow{3}{*}{585} & \multirow{3}{*}{1910} & STEM\\
\cellcolor{gray!20} Development&&&&&\&\\ 
\cellcolor{gray!20}Fundamentals&&&&&Non-MOOC\\
\hline
\end{tabular}
}
\label{tab:details}
\end{table}
\vspace{-0.1cm}

% \vspace{-.5mm}
\subsubsection{Temporal Representation of SLNs}
The MOOC datasets (i.e., \textit{`algo', `comp', `shake', and `ml'}) are static longitudinal data logs capturing interactions between online student pairs, with links formed cumulatively over a certain period. However, since explicit timeline information was not provided, we imposed a temporal structure by assuming the absence of links among certain pairs at an earlier time, thereby reflecting the potential for those links to form later. %creating two versions of our dataset. The first version is identical to the original dataset, representing data at time step $t$. 
To emulate this temporal evolution, we created a modified version of the original datasets representing time step $t-1$ by randomly selecting 20\% of the data (i.e., student pairs) and removing any existing links between the corresponding student pairs. The assumption is that the removed links would emerge at time step $t$, as reflected in the original datasets. Using this setup, we employed the data at time step $t-1$ to predict link formations at time step $t$. 
% Followed by the computation of topological values of the node pairs for the modified SLNs based on their updated link status, we transformed skewed feature values and finalized the datasets (detailed feature descriptions can be found in Section~\ref{sec:feature}). 
Nonetheless, the \textit{`csc'} dataset contains student interaction data recorded continuously throughout the semester. For this dataset, we captured the first instance of link formation between student pairs. Subsequently, we used the SLN topology at week 10 (i.e., time $t-1$) to predict the SLN topology --- i.e., the links among students --- in weeks 11, 12, 13, and 14 (i.e., time $t$). 
To evaluate our prediction performance, we partitioned the data into training and test sets, allocating 80\% of the student pairs to the training set (i.e., 80\% of student pairs at time $t-1$) and the remaining 20\% to the test set.

% \vspace{-1.85mm}
\subsubsection{Feature Description}\label{sec:feature}
To extract the features from data points (i.e., student pairs), given the fact that SLNs resemble graph structures, we employ a set of graph-theoretic features used in various link prediction tasks \cite{hasan2011survey}. In a nutshell, these pairwise features measure the similarity between node pairs based either on the similarity of their neighborhoods in the graph 
or on the connectivity paths between them~\cite{lu2011link,liben2003link,yun2021neo,ma2024mixture,bojanowski2020proximity}. 
% They are domain-independent, meaning they do not require any specific graph structure and thus can be computed without domain-specific knowledge. 
% Given that we only had information about pairs of nodes and the presence or absence of direct paths between them, the simplest approach to link prediction in our case was the proximity-based algorithm \cite{lu2011link}, where each pair of nodes, $u$ and $v$, are assigned a score that represents their proximity or similarity \cite{liben2003link}. Proximity-based features quantify the closeness or relatedness of nodes %based on their positions and connections in the graph. These features assess how easily one node can reach another and the strength of their connections, 
% often relying on shared neighbors. On that note, we considered neighborhood overlap-based metrics \cite{yun2021neo, ma2024mixture}, which assess the similarity or overlap of node neighborhoods, and based on graph topology,
In particular, for each of the five SLNs, we computed six features --- enumerated from \textbf{(Feature I)} to \textbf{(Feature VI)} below --- for every learner node pair \((u, v)\), where \(u \neq v\),  corresponding to nodes/students $S_u$ and $S_v$. %We used Python package \textit{networkx}. For neighbor-based pairwise calculations, 
Henceforth, we use $\Gamma_u$ to denote the set of neighboring nodes of $S_u$, i.e., the nodes that have a link/edge with $S_u$. Consequently, \(|\Gamma_u|\) represents the \textit{degree} of \(S_u\), i.e., the number of edges connected to node $S_u$.

% Here, \(N(G)\) is the set of nodes in the SLN \(G\), \(\Gamma_u \subseteq N(G)\) denotes the set of neighbors of \(u\), \(\Gamma_n\) is the set of common neighbors of \(u\) and \(v\), and \(|\Gamma_u|\) is the degree of \(u\), i.e., the number of edges that \(u\) has.

%, the neighborhood-based features measure the ``similarity'' of \(u\) and \(v\)'s neighborhoods . 
%The binary output (0/1) was decided regarding the existence of a link. %We did not have related metadata to compute path- and post-based \cite{hasan2011survey, liben2003link} features. 
% The six features are described below:
% \vspace{-0.2cm}
% \begin{description}

\vspace{-2.5mm}
    \textbf{(Feature I) Jaccard Coefficient \cite{jaccard1912distribution}:} 
    This metric quantifies the similarity between two nodes $S_u$ and $S_v$ through a ratio, capturing the proportion of their shared neighbors as follows:
    \vspace{-3.0mm}
    \begin{equation}
        \mathcal{J}_{uv} = \frac{|\Gamma_u \cap \Gamma_v|}{|\Gamma_u \cup \Gamma_v|},
        \label{eq:jaccard}
        % \vspace{-.15mm}
    \end{equation}
    where the numerator represents the number of common neighbors between nodes \(S_u\) and \(S_v\), while the denominator indicates the total number of distinct neighbors across both nodes. A higher Jaccard Coefficient value between two students implies a greater proportion of shared connections between them, suggesting they belong to closer social circles. %many neighbors, and this strong overlap in their neighborhoods can increase their potential to be connected. %This metric is widely used to predict the likelihood of a connection between two nodes based on their shared neighbors.

\vspace{-2.5mm}
    \textbf{(Feature II) Adamic-Adar Index \cite{adamic2003friends}:}
    This metric quantifies the likelihood of a connection between two nodes $S_u$ and $S_v$ through the inverse logarithm of the degrees of their shared neighbors as follows:
    \vspace{-2.2mm}
    \begin{equation}
        \mathcal{A}_{uv} = \sum_{S_n \in \Gamma_u \cap \Gamma_v} \frac{1}{\log|\Gamma_n|},
        \label{eq:aai}
        %\vspace{-.5mm}
    \end{equation}
    where the summation is taken over all common neighbors of \(S_u\) and \(S_v\). In essence, the inverse of \( \log|\Gamma_n| \) assigns higher weights to neighbors with fewer connections,
    % A low-degree mutual neighbor has more influence in an increased Adamic-Adar value between the two nodes than a high-degree one. 
    implying that two students sharing a rarely active peer as a common neighbor are more likely to be connected than when they share a highly active peer. %Higher values of the Adamic-Adar Index indicate a higher likelihood of the existence of a link between the two nodes, suggesting that they share many common neighbors, especially neighbors with few other connections. This metric gives more importance to neighbors who are not well-connected themselves, emphasizing the significance of uncommon or rare connections in the network.

        \vspace{-2.5mm}
    \textbf{(Feature III) Resource Allocation Index \cite{zhou2009predicting}:} 
    This metric, similar to the Adamic–Adar Index \eqref{eq:aai}, quantifies the likelihood of a connection between two nodes based on the inverse of the degrees of their common neighbors as follows:
    \vspace{-1mm}
    \begin{equation}
       \mathcal{R}_{uv} = \sum_{S_n \in \Gamma_u \cap \Gamma_v} \frac{1}{|\Gamma_n|}.
       %\vspace{-.5mm}
    \end{equation}
     % Like the Adamic-Adar Index, this metric  indicates a stronger potential for two nodes connecting for shared neighbors with fewer connections.
     A higher value of the Resource Allocation Index between two nodes/students suggests that their common neighbors have fewer connections --- meaning these neighbors are selective in their interactions and are not connected to the majority of students. This selectivity implies a stronger likelihood of a direct connection forming between the two students. %This measure is particularly useful in sparse networks or when studying social ties. 

    \vspace{-2.5mm}
    \textbf{(Feature IV) Preferential Attachment Score \cite{barabasi1999emergence}:} 
    Unlike previous metrics that focus on node neighborhoods, this metric directly assesses the similarity between two nodes 
$S_u$ and $S_v$ based on their degrees and is defined as follows:
\vspace{-.5mm}
    \begin{equation}
        \mathcal{P}_{uv} = |\Gamma_u| \cdot |\Gamma_v|.
        %\vspace{-.5mm}
    \end{equation} 
    % A higher Preferential Attachment Score between two students/nodes suggests a higher likelihood that they will form a link if not already connected. This means in classrooms, students with many existing connections are more likely to acquire additional connections, and students with fewer neighbors will tend to form a lesser number of connections.
    A higher Preferential Attachment Score between two students/nodes indicates a greater likelihood of a link forming between them if one does not already exist. In the context of classrooms, this suggests that students with many existing connections are more likely to develop additional connections.

   \vspace{-2.5mm}
    \textbf{(Feature V) Cosine Similarity \cite{salton1975vector}:} 
    Also known as the Salton Index, the Cosine Similarity measures the similarity between the two nodes $S_u$ and $S_v$ by computing a value that resembles the cosine of the angle between two vectors as follows:
    \vspace{-.5mm}
    \begin{equation}
        \mathcal{C}_{uv} = \frac{|\Gamma_u \cap \Gamma_v|}{\sqrt{|\Gamma_u| \cdot |\Gamma_v|}},
        \label{eq:cosine}
        %\vspace{-.5mm}
    \end{equation}
    where the numerator represents the number of common neighbors between nodes/students \(S_u\) and \(S_v\), while the denominator represents the geometric mean of their individual degrees.  Unlike the Jaccard Coefficient  \eqref{eq:jaccard}, which considers the ratio of shared to total neighbors, the Cosine Similarity focuses on the alignment of the nodes' neighborhoods. A higher Cosine Similarity score between a student pair suggests that their neighborhoods are more aligned, implying stronger similarity and a higher likelihood of connectivity between them.

    \vspace{-2.5mm}
  \textbf{(Feature VI) Dice Similarity \cite{dice1945measures}:} 
    While the Jaccard Coefficient \eqref{eq:jaccard} excludes duplicates when counting the neighbors of two nodes, this metric counts such duplicates, defined as
    \vspace{-.5mm}
    \begin{equation}
        \mathcal{D}_{uv} = \frac{2 \cdot |\Gamma_u \cap \Gamma_v|}{|\Gamma_u| + |\Gamma_v|},
       % \vspace{-.5mm}
    \end{equation}
    where the numerator represents twice the number of common neighbors between nodes \(S_u\) and \(S_v\), and the denominator reflects the total number of neighbors of both nodes.  By doubling the weight of shared neighbors, this metric assigns greater importance to common connections compared to the Cosine Similarity \eqref{eq:cosine}.
A higher Dice Similarity between a student pair indicates a significant overlap in their social circles, suggesting a stronger likelihood of forming a connection. 

% Conversely, a lower score implies that either student has a larger, less-overlapping set of neighbors, thereby reducing the probability of their connection. %, large proportion of each node's neighbors are shared, which can lead to their %could signify a close relationship or shared involvement in specific clusters or activities within the network.

    % \item[(vii) Neighborhood Overlap Similarity :]
    % This metric captures the proportion of overlap in the neighborhoods of two nodes, as follows:
    % \begin{equation}
    %     \mathcal{O}_{uv} = \frac{|\Gamma_u \cap \Gamma_v|}{\min(|\Gamma_u|, |\Gamma_v|)},
    % \end{equation}
    % % which is similar to Equation \ref{eq:jaccard}, with the key difference being that the denominator considers the smaller neighborhood size instead of the union. A higher Neighborhood Overlap Similarity indicates that two students have a strong overlap in their neighborhoods, increasing their potential to be connected.  
% % \end{description}

\vspace{-2mm}
\subsection{Centralized Learning Architecture}\label{sec:3-b}
% We developed a centralized model for the link prediction task as a baseline for our distributed FL model. A centralized model refers to one single model trained on all available datasets. In our study, the centralized model was trained and tested with the aggregated data from five different classrooms. The purpose of this model is to compare the effectiveness of the distributed FL model. However, the centralized model strictly hinders the privacy of the local datasets/classrooms as they are all shared directly during the model training. We implemented a convolutional neural network (CNN) model for the prediction task in alignment with Sahay et al.'s best model \cite{sahay2023predicting}. For hyperparameter tuning, the learning rate was 0.001, the batch size set was 128, 256, 512, and 1024, and the set of epochs was 50, 100, and 200. %We noted that a higher learning rate made the training faster but the learning unstable, and the more the epochs were, the better it was for a slower but more stable convergence. 
% We checked for overfitting by comparing training loss and validation loss and tuned the model hyperparameters with a 5-fold cross-validation approach.  

Using the above features for every student pair in each of the five SLNs of our interest described in Table~\ref{tab:details}, we developed a centralized model as a baseline for evaluating the performance of our FL models in the link prediction task. A centralized model refers to a single model trained on the complete set of available data, which is trained and tested using the aggregated data from five distinct classrooms. It is important to note that \textit{this centralized model compromises the privacy of local datasets}, as data from individual classrooms are shared during the training process.
Our central model is a convolutional neural network (CNN), following the architecture used in \cite{sahay2023predicting}, trained using the stochastic gradient descent (SGD) method. To optimize its performance, we conducted hyperparameter tuning with the following configurations: the learning rate choices of \{0.1, 0.01, 0.001\}, mini-batch size choices of \{64, 128, 256, 512, 1024\}, and epoch counts of \{50, 100, 200\}. We monitored for overfitting by comparing the training and validation losses and used a 5-fold cross-validation to tune the hyperparameters.

\vspace{-2mm}
\subsection{Federated Learning (FL) Architecture}\label{sec:3-c}
\begin{figure}[t]
    \centering
    \includegraphics[width=0.29\textwidth, alt={FL's distributed training architecture illustration}]{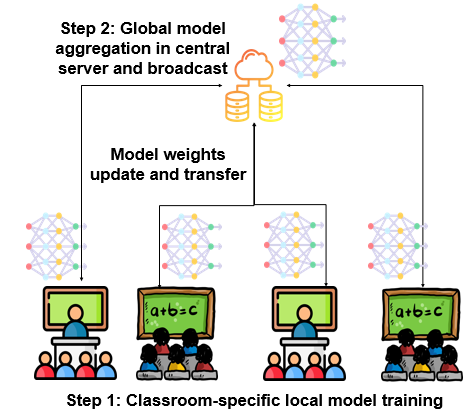}
    \caption{A schematic of FL's distributed training architecture involving the data of multiple classrooms.}
    \label{fig:schematic}
\end{figure}

Our (vanilla) FL method is built upon the FedAvg architecture \cite{mcmahan2017communication}, utilizing the same CNN architecture employed in the centralized model training. In this setup, each SLN is assumed to reside in a separate storage space with local computing capabilities (e.g., a local server, hereafter referred to as a \textit{client}). A central server --- either a cloud server or another local server --- coordinates the distributed model training process. In a nutshell, as depicted in Figure \ref{fig:schematic}, the training procedure consists of two main phases: \textit{local model training} and \textit{global model aggregation}. During local training, each client first synchronizes its local model with the latest model received from the central server, independently trains its local model on its dataset, and subsequently sends the model parameters to the central server. The central server then performs a global model update by aggregating the received parameters using weighted averaging, thereby forming an updated global model that is redistributed to the clients for the next training round.
In the following, we formally describe these processes.

Let $\mathbf{w}^{(k)}$ denote the global model parameters maintained by the central server at the end of the $k$-th global aggregation round. These parameters are broadcast to clients at the beginning of the subsequent round $k+1$. The model training process initiates at $k=0$, with the central server generating the initial global model $\mathbf{w}^{(0)}$ (e.g., through random initialization) and distributing it to all clients. We next describe the procedure through which clients utilize $\mathbf{w}^{(k)}$ during the $k$-th global round to derive the updated global model $\mathbf{w}^{(k+1)}$.
% In our FL model, a global model was initiated on a central server. This global model was then sent to participating local clients (local servers) that had their own datasets from the local classrooms. 

Let $\mathcal{I}$ denote the set of clients in the system, and let $\mathcal{D}_i$ with size $|\mathcal{D}_i|$ represent the dataset of client \( i \in \mathcal{I} \), corresponding to a server with the data of one of the SLNs. The loss function of client $i$ for an arbitrary model parameter $\mathbf{w}$ is then given by $
\mathcal{L}_i(\mathbf{w}) = \frac{1}{|\mathcal{D}_i|} \sum_{d \in \mathcal{D}_i} \ell(\mathbf{w}; d),
$ 
where $\ell(\mathbf{w}; d)$ is the ML loss function that quantifies the prediction performance of the model parameter $\mathbf{w}$ for data point $d$ (e.g., cross-entropy loss).
Upon receiving $\mathbf{w}^{(k)}$, each client \( i \) initializes its local model as follows $
\mathbf{w}_i^{(k),0} \leftarrow \mathbf{w}^{(k)},$
and subsequently updates its local model through a series of SGD iterations, indexed by $e$, on its local dataset. Let $e_{\text{max}}$ denote the number of local SGD iterations performed by the clients. The evolution of the local model of client $i$ across the SGD iterations is given by
\vspace{-.2mm}
\begin{equation}\label{eq:SGDs}
\mathbf{w}_i^{(k),e} = \mathbf{w}_i^{(k),e-1} - \eta \, \widetilde{\nabla} \mathcal{L}_i\big(\mathbf{w}_i^{(k),e-1}\big),
%\vspace{-.2mm}
\end{equation}
where $e \in \{1, \cdots, e_{\text{max}}\}$ indicates the SGD iteration count, $\eta$ denotes the learning rate, and $\widetilde{\nabla} \mathcal{L}_i(\mathbf{w}_i^{(k),e-1})$ is the stochastic approximation of the gradient of the local loss function, computed as $
\widetilde{\nabla} \mathcal{L}_i(\mathbf{w}_i^{(k),e-1}) = \frac{1}{|\mathcal{B}_i^{(k),e}|} \sum_{d \in \mathcal{B}_i^{(k),e}} \nabla \ell(\mathbf{w}_i^{(k),e-1}; d),
$ where $\mathcal{B}_i^{(k),e} \subseteq \mathcal{D}_i$ denotes a randomly selected mini-batch of data from the local dataset, and $|\mathcal{B}_i^{(k),e}|$ is its size.
After completing the local model training, each client sends its latest local model parameters $\mathbf{w}_i^{(k),e_{\text{max}}}$ to the central server. It is worth noting that while the structure of the local models is consistent with that of the global model (i.e., all models share the same CNN architecture as in centralized training), the key difference lies in their trained parameters, which reflect the variations in local datasets and the differing patterns of student interactions they represent.

After the reception of the local models $\{\mathbf{w}_i^{(k),e_{\text{max}}}\}_{i\in\mathcal{I}}$, the central server updates the global model as follows:
\vspace{-.2mm}
\begin{equation}\label{eq:agg}
    \mathbf{w}^{(k+1)} = \sum_{i\in\mathcal{I}} \frac{\big|\mathcal{D}_i\big|\mathbf{w}_i^{(k),e_{\text{max}}}}{|\mathcal{D}|} ,
    %\vspace{-.2mm}
\end{equation}
where $|\mathcal{D}|=\sum_{i\in\mathcal{I}}|\mathcal{D}_i|$ denotes the cumulative size of datasets distributed across the clients. The new global model $\mathbf{w}^{(k+1)}$ is then broadcast across the client to synchronize their local models and initiate the next round of local model training and global model aggregations.

In our implementation, the hyperparameters of FL we optimized through a grid search as follows: for each client, the mini-batch size $B=|\mathcal{B}_i^{(k),e}|$, $\forall k,e$, choices of \{128, 256\} and the number of SGD steps $e_{\text{max}}$ choices of \{100, 200\}. 
\vspace{-1mm}
\subsection{Personalization of FL Models}\label{sec:3-d}
% Computing maximum differences between the cumulative distribution functions (CDFs) of the features across the five datasets, Kolmogorov-Smirnov (KS) test showed that they have significantly different distributions, indicating clients' datasets are non-independent and identically distributed (non-IID). As each client's dataset has a unique feature distribution, training a single global FL model and use it for all the clients may lead to a sub-optimal performance. Therefore, to better capture each SLN's data nuances, we applied three personalized FL (pFL) methods which are sorted with respect to their implementation complextity detailed below where --- through some changes during the training and post-training phases of FL --- the global model gets tailored to the individual clients' datasets.

Computing the maximum differences between the cumulative distribution functions (CDFs) of the features across the five datasets, the Kolmogorov-Smirnov (KS) test confirmed that they follow different distributions, implying that the clients' datasets are non-independent and identically distributed (non-IID).
Since each client's dataset exhibits a unique feature distribution, training a single global FL model and applying it uniformly across all clients, as done in vanilla FedAvg, may lead to suboptimal performance. Subsequently, to better capture the nuances of each SLN's data, we implemented three personalized FL (pFL) methods, ranked by their implementation complexity, as detailed below. These methods introduce modifications during the training and post-training phases of FL, allowing the global model to be tailored to individual clients' datasets.

\vspace{-1mm}
\textbf{1) FedAvg followed by Local Fine-Tuning (FedAvg+FT):}  
In this pFL method, a post-training fine-tuning process is applied to the global model obtained via FedAvg. Let $\mathbf{w}^{(K)}$ denote the final global model obtained at the end of the $K$-th global aggregation round of vanilla FedAvg, marking the conclusion of the training phase. 
In this approach, the server broadcasts $\mathbf{w}^{(K)}$ to all clients. Each client then adapts the global model to its local dataset by performing an additional epoch of local training. Notably, conducting more epochs of local fine-tuning often causes the model to forget the globally shared data patterns across clients, potentially reducing generalization. After this fine-tuning step, the client uses the adapted local model for inference on its dataset.

%These methods prove more useful in such cases when data heterogeneity is a concern. %While finetuning occurs independently for local clients only after the FL training is complete, this personalization approach adapts local data as training happens.  
% \vspace{-.1cm}
% \begin{description}

% \ali{TO BE INTEGRATED HERE AND EDITED: After several iterations of global aggregation, the global model was finalized and distributed to each local node representing individual classrooms. At this point, personalization occurred: the global model was copied to each local model and finetuned using only the local classroom dataset.
% The implication of this personalization process is significant in a classroom setting. By finetuning the global model with classroom-specific data, we adapt the model to the distinct learning patterns, interaction styles, and other unique factors in each class. This allows for a more accurate and context-sensitive model for each classroom, improving its ability to provide more relevant predictions based on the individual learning dynamics of students from distinct classrooms/courses. While the global model captures broader trends across all classrooms, this personalization step enhances the model's effectiveness in addressing the specific needs of each group of students.
% }

\vspace{-1mm}
   \textbf{2) Personalized FedAvg via Meta-Learning using Hessian-Free Approximation Method (PerFedAvg-HF)~\cite{fallah2020personalized}:}  
This pFL method adopts the same aggregation rule as FedAvg (see Sec.~\ref{sec:3-c}) but introduces modifications to both its training and post-training phases. The primary goal of PerFedAvg-HF is to obtain a global model during the training phase that is \textit{adaptable} to each client’s local dataset at the end of training.
To achieve this adaptability, instead of performing standard SGD iterations over the original loss function $\mathcal{L}_i(\mathbf{w})$ (as in vanilla FedAvg), each client $i$ executes SGD iterations over a modified loss function defined as:
$\mathcal{L}_i\big(\mathbf{w} - \eta~\nabla\mathcal{L}_i(\mathbf{w})\big),
$ where $\eta$ denotes the learning rate. This modification encourages the global model to be more amenable to local adaptation: after one epoch of local fine-tuning, the model can be efficiently adapted to client $i$'s dataset, captured by the update $\mathbf{w} - \eta~\nabla\mathcal{L}_i(\mathbf{w})$.
However, performing SGD over this modified loss function typically requires computing the Hessian matrix, which is computationally intensive. To address this issue, we employ the Hessian-free approximation method proposed in~\cite{fallah2020personalized}.
Similar to FedAvg, after every $e_{\text{max}}$ local SGD iterations over the modified loss function, clients send their updated local models to the server for aggregation according to the same rule as in~\eqref{eq:agg}. Also, following the conclusion of the training phase, the final global model is broadcast back to the clients. Afterward, each client performs one additional epoch of fine-tuning on its local dataset, similar to the process described in the FedAvg+FT method.

    \vspace{-1mm}
    \textbf{3)  Federated Adaptive Local Aggregation (FedALA)~\cite{zhang2023fedala}:}
    The local model training and global aggregation processes of FedALA are similar to vanilla FedAvg (see Sec.~\ref{sec:3-c}); however, instead of directly overwriting/synchronizing the local model with the global model after the reception of the global model at each client, FedALA employs an adaptive local aggregation (ALA) module to element-wise aggregate the received global model from the server $\mathbf{w}^{(k)}$ and the latest local model $\mathbf{w}_i^{(k-1),e_{\text{max}}}$ at each client $i$. Specifically, instead of initializing the local model as $\mathbf{w}_i^{(k),0} \leftarrow \mathbf{w}^{(k)}$ as in other FL methods, in FedALA the local model is initialized as
$
\mathbf{w}_i^{(k),0} \leftarrow \mathbf{w}_i^{(k-1),e_{\text{max}}} + (\mathbf{w}^{(k)} - \mathbf{w}_i^{(k-1),e_{\text{max}}}) \odot \mathbf{W}_i,
$
where $\mathbf{W}_i$ is a learnable, element-wise aggregation weight matrix constrained to $[0,1]$ via a clipping function to ensure stability and $\odot$ denotes the Hadamard product. This adaptive weighting enables FedALA to capture desired information from the global model for each client while mitigating the impact of undesired information (i.e., those that do not relate to the client's local data), thus enhancing local model personalization. FedALA further considers the fact that lower layers in deep neural networks typically extract fundamental features that are broadly applicable across clients, whereas higher layers capture more task- or client-specific features. Subsequently, in this method, the ALA module is applied only to the top $p$ layers ($p$ is a hyperparameter) during the local model synchronization, allowing clients to retain shared generic representations from the global model in the lower layers without modification, ensuring better generalizability of the local models. Further, FedALA introduces a hyperparameter $\zeta$, determining the fraction of the local dataset utilized to learn the ALA weights $\mathbf{W}_i$ for each client $i$. The hyperparameters of FedALA were optimized through a grid search: for $p$, the choices were \{1,2\}, while for $\zeta$, the choices covered interval $[5,100]$ with increments of $5$.

\vspace{-2mm}
\subsection{Fairness Measures}\label{sec:3-e}
We consider two widely adopted fairness measures in the literature: \textit{equalized odds} and \textit{equalized opportunity}~\cite{hardt2016equality}. Both measures are built upon quantifiable metrics --- namely the true positive rate (TPR), which assesses the model's ability to correctly identify students who form links, and the false positive rate (FPR), which evaluates how often students not connected are incorrectly classified as link-forming.
Let $d$ denote an arbitrary data point (i.e., a student pair), $\widehat{Y}(d)$ the predicted label, and ${Y}(d)$ the true label (i.e., $0$ or $1$ depending on the existence of an edge) of the data point. In this study, our goal is to assess the fairness of the model across various classrooms/SLNs (e.g., ensuring the FL global model treats students from different classrooms equally well). In this context, the \textit{equalized odds} criterion is satisfied when both of the following probabilistic conditions hold for every pair of SLNs/clients $i, i' \in \mathcal{I}$. \textit{(Condition 1) Equality of TPRs:}
$
\mathbb{P}\big(\widehat{Y}(d) = 1 \big| Y(d) = 1, d \in \mathcal{D}_i\big) = \mathbb{P}\big(\widehat{Y}(d) = 1 \big| Y(d) = 1, d \in \mathcal{D}_{i'}\big)
$, \textit{(Condition 2) Equality of FPRs:} $
\mathbb{P}\big(\widehat{Y}(d) = 1 \big|  Y(d) = 0, d \in \mathcal{D}_i\big) = \mathbb{P}\big(\widehat{Y}(d) = 1 \big| Y(d) = 0, d \in \mathcal{D}_{i'}\big)$.
In essence, equalized odds is achieved when both the TPR (i.e., Condition 1) and FPR (i.e., Condition 2) are equal across clients/SLNs. 
Further, \textit{equal opportunity} is a relaxed version of equalized odds, requiring only the equality of TPRs (i.e., the satisfaction of Condition 1) across clients.
In our subsequent simulations, \textit{we evaluate the fairness of the models by examining how closely their TPRs and FPRs adhere to these fairness criteria}, thereby assessing the extent to which the models satisfy equalized odds and equal opportunity.

\subsection{Model Explainability}\label{sec:3-f}
To explain the model's decision-making process and interpret the influence of individual features, we use SHapley Additive exPlanations (SHAP), an XAI method grounded in game theory~\cite{lundberg2017unified}. SHAP quantifies the contribution of individual features to a model's predictions by calculating \textit{Shapley values} for each feature, providing both global explanations (overall feature importance) and local explanations (feature impact on individual predictions). Formally, SHAP measures
% \begin{equation}
%     \phi_i = \sum_{S \subseteq N \setminus \{i\}} \frac{|S|!(\big|N\big| - \big|S\big| - 1)!}{\big|N\big|!} \left[ v(S \cup \{i\}) - v(S) \right],
% \end{equation}
\begin{equation} \label{eq:SHAP}
\hspace{-6mm}
    \varphi_{_f} = \sum_{\mathcal{F} \subseteq \mathcal{N} \setminus \{f\}} \hspace{-2mm} \frac{\big|\mathcal{F}\big|! (n - \big|\mathcal{F}\big| - 1)!}{n!} \Big( v(\mathcal{F} \cup \{f\}) - v(\mathcal{F}) \Big),\hspace{-3mm}
\end{equation}
for every feature \( f \) (i.e., the six features discussed in Sec.~\ref{sec:feature}) where \( \varphi_{_f} \) is the Shapley value for feature \( f \), indicating its contribution to the prediction.
The summation in~\eqref{eq:SHAP} iterates over all possible subsets \( \mathcal{F} \) of features excluding \( f \), where \( \mathcal{N}  \) is the set of all features and $n$ denotes the number of features ($n=6$ in our case). 
The term \( \big|\mathcal{F}\big|! (n - \big|\mathcal{F}\big| - 1)! / n! \) is a weighting factor that ensures each subset \( \mathcal{F} \) is considered fairly by accounting for all permutations in which feature \( f \) could be added. The function \( v(\mathcal{F}) \) represents the \textit{value function} --- measuring the model's prediction --- of the subset \( \mathcal{F} \), and the term \( v(\mathcal{F} \cup \{f\}) - v(\mathcal{F}) \) measures the marginal contribution of feature \( f \) to the prediction performance when added to the subset \( \mathcal{F} \).
In essence, by examining how the model's predictions change when features are omitted, SHAP identifies the positive or negative influence of each feature. We will utilize SHAP to produce visualizations that illustrate the importance of individual features across different classrooms/SLNs and their impact on the likelihood of link formation between students.

% \ali{EDITED UP TO HERE! WILL BE BACK!}
%and their implications on the links formed among students. %The generated SHAP plots provide an explainable, feature-level breakdown of the model’s predictions, offering actionable insights into how specific features influence the prediction outcomes. 
%SHAP calculates Shapley values for each feature and instance, quantifying the contribution and importance of each feature to both the overall predictive model (global explanations) and individual predictions (local explanations). This model-agnostic framework splits the variability of predictions across the features used in the prediction model.
%In SHAP, the Shapley values represent the presence of each covariate in the model’s predictions as a linear combination of the predictor variables. By examining how the prediction changes when a feature is withheld, SHAP determines the positive or negative effect of each feature on the final predictions. This enables us to measure the contribution of each feature behind the predictive model. 

\vspace{-1mm}
\section{Results \& Discussions}
In this section, we report performance evaluation~(Sec.~\ref{sec:4-a}) and fairness comparison~(Sec.~\ref{sec:4-b}) to demonstrate the effectiveness of personalized/fine-tuned FL models over the centrally trained model and discuss insights obtained from applying XAI methods~(Sec.~\ref{sec:4-c}). Finally, we offer implications of this study for the educators and institutions~(Sec.~\ref{sec:4-d}).
%global E=350, local E=15
% \vspace{-.1cm}
\begin{table*}[ht]
\centering
\caption{Comparison of performance ($\pm$ standard deviation) between implemented methods shows that the personalized federated model FedALA outperformed the centralized model for all datasets. Metric values in \textbf{bold} denote the best results.\vspace{-.3cm}}%, confirmed by a significance test.}
{\small
\begin{tabular}{|>{\centering\arraybackslash}m{.8cm}||>{\centering\arraybackslash}m{1.6cm}|>{\centering\arraybackslash}m{1.8cm}|>{\centering\arraybackslash}m{2.5cm}|>{\centering\arraybackslash}m{2cm}|>{\centering\arraybackslash}m{2.6cm}|>{\centering\arraybackslash}m{3.2cm}|}
% \begin{tabular}{|c||c|c|c|c|c|c|}
\hline
\multirow{3}{*}{\textbf{Course}} & \multirow{3}{*}{\textbf{Metrics}} & \textbf{Centralized} & \textbf{FedAvg~\cite{mcmahan2017communication}} ($\eta$= & \textbf{FedAvg+FT} & \textbf{PerFedAvg-HF~\cite{fallah2020personalized}} &\textbf{FedALA~\cite{zhang2023fedala}} ($\eta$=0.01,\\ 
& & ($\eta$=0.001, E=& 0.001, K=30, & ($\eta$=0.0001, B=& ($\eta$=0.01, B=256,%$\beta$=0.00045, &=0.01$\rightarrow0.001$, T=30,\\
&K=30, $e_{\text{max}}$=100,\\
&& 200, B=256)& $e_{\text{max}}$=200, B=256)& 64, $e_{\text{max}}$=200) & $e_{\text{max}}$=350, K=15) & B=128, p=2, $\zeta$=80)\\ 
%&&& && B=1024)& B=256)&E=5, B=8)\\
 \hline
 \hline%\hline
 %\rowcolor{green!20} % First row color
\cellcolor{green!20} & Accuracy(\%) & $93.62\pm0.34$ & 
$93.26\pm0.26$ & $93.88\pm0.01$ & $93.7\pm0.03$ & $\textbf{94.01}\pm\textbf{0.14}$\\ 
\cellcolor{green!20} algo &Loss & $0.18\pm0.01$ & $0.2\pm0.01$ & $0.17\pm0.01$ &$0.2\pm0.002$ &$\textbf{0.17}\pm\textbf{0.002}$\\ 
\cellcolor{green!20} &AUC & $0.9\pm0.002$ & $0.9\pm0.001$ & $0.89\pm0.01$ &$0.9\pm0.01$ & $\textbf{0.91}\pm\textbf{0.001}$\\ 
\hline
%\rowcolor{red!20} % First row color
\cellcolor{red!20} & Accuracy(\%) & $92.26\pm0.19$ & 
$92.22\pm0.06$ & $92.65\pm0.15$ & $92.00\pm0.37$ & $\textbf{92.76}\pm\textbf{0.04}$\\ 
\cellcolor{red!20} comp &Loss & $0.2\pm0.01$ & $0.22\pm0.2$ & $0.18\pm0.003$ &$0.21\pm0.001$ &$\textbf{0.18}\pm\textbf{0.002}$\\ 
\cellcolor{red!20} &AUC & $0.86\pm0.001$ & $0.86\pm0.01$ & $0.85\pm0.002$ &$0.84\pm0.001$ & $\textbf{0.86}\pm\textbf{0.01}$\\ 
\hline
\cellcolor{blue!20} & Accuracy(\%) & $90.52\pm0.18$ & 
$90.27\pm0.08$ & $91.2\pm0.05$ & $90.98\pm0.13$ & $\textbf{91.21}\pm\textbf{0.05}$\\ 
\cellcolor{blue!20} shake &Loss & $0.22\pm0.003$ & $0.23\pm0.01$ & $0.2\pm0.01$ &$0.23\pm0.001$ &$\textbf{0.2}\pm\textbf{0.001}$\\ 
\cellcolor{blue!20} &AUC & $0.84\pm0.01$& $0.84\pm0.01$ & $0.8\pm0.01$ &$0.83\pm0.001$ & $\textbf{0.84}\pm\textbf{0.01}$\\ 
\hline
\cellcolor{orange!20} & Accuracy(\%) & $86.31\pm0.15$ & 
$84.93\pm0.17$ & $86.7\pm0.34$ & $86.93\pm0.31$ & $\textbf{87.2}\pm\textbf{0.32}$\\ 
\cellcolor{orange!20} ml &Loss & $0.33\pm0.002$ & $0.37\pm0.01$ & $0.32\pm0.01$ &$0.36\pm0.002$ &$\textbf{0.32}\pm\textbf{0.01}$\\ 
\cellcolor{orange!20} &AUC & $0.85\pm0.002$ & $0.82\pm0.01$ & $0.86\pm0.001$ &$0.86\pm0.001$ & $\textbf{0.86}\pm\textbf{0.01}$\\ 
\hline
\cellcolor{gray!20} & Accuracy(\%) & $76.57\pm0.59$ & 
$75.52\pm1.77$ & $79.71\pm0.004$ & $81.48\pm0.15$ & $\textbf{82.11}\pm\textbf{2.51}$\\ 
\cellcolor{gray!20} csc &Loss & $0.62\pm0.05$ & $0.56\pm0.01$ & $0.44\pm0.02$ &$0.48\pm0.03$ &$\textbf{0.41}\pm\textbf{0.01}$\\ 
\cellcolor{gray!20} &AUC & $0.58\pm0.01$ & $0.61\pm0.01$ & $0.7\pm0.02$ &$0.76\pm0.01$ & $\textbf{0.77}\pm\textbf{0.01}$\\ 
\hline 
\end{tabular}
}
\label{tab:res_baseline}
\end{table*}
\vspace{-.1cm}

\vspace{5mm}
\subsection{Link Formation Prediction}\label{sec:4-a}
%First paragraph (may be split into two paragraphs if needed): Explain what we present here (i.e., FL results) and explain each of the considered baselines. Also explain how the baselines are trained. 
%In Table~\ref{tab:res_baseline}, we present the performance of our centralized baseline model for link prediction, alongside the results of FedAvg and the pFL variations. As discussed in Sec.~\ref{sec:3-b}, the centralized training model serves as the \textit{ideal baseline} since it does not adhere to privacy restrictions.
%In the table, the top row lists the various methods along with their corresponding hyperparameters, where $E$ denotes the number of training epochs in the centralized training scenario, while $K$ represents the number of global aggregation rounds in the FL settings and the remaining notations are defined in Sec.~\ref{sec:3-c}. All the models reached convergence under the specified settings.
\vspace{1mm}
Here, we present the performance evaluation of our proposed link prediction methodology. We compare our approach to a centralized baseline model for link prediction, along with FedAvg and variations of pFL. As discussed in Sec.~\ref{sec:3-b}, the centralized training model serves as the \textit{ideal baseline} as it has access to the pool of data of all the SLNs and does not adhere to privacy restrictions.

To quantify the performance, we calculated the accuracy, loss, and area under curve (AUC) of the models on each SLN's test data and reported them in Table~\ref{tab:res_baseline}. AUC measures the probability that a randomly chosen positive instance (i.e., a connected student pair) is ranked higher than a randomly chosen negative instance (i.e., a non-connected pair) by the model. An AUC of 0.5 indicates no discriminative power (equivalent to random guessing), while an AUC of 1.0 signifies perfect class separation. As seen in Sec.~\ref{sec:3-a}, SLNs are often sparse graphs \cite{xu2018many} showing class imbalance (i.e., most of the student pairs are not connected); hence, AUC unveils how well classes are separated \cite{saito2015precision}. 

We present our results in Table \ref{tab:res_baseline}, where the top row lists the various methods along with their corresponding hyperparameters, where $E$ denotes the number of training epochs in the centralized training scenario, while $K$ represents the number of global aggregation rounds in the FL settings and the remaining notations are defined in Sec.~\ref{sec:3-c}. All the models reached convergence under the specified settings.

% While accuracy values change across runs, loss stays more or less stable, suggesting the training process is stable, and the optimizer reaches a similar state every time. 

%AUC also does not vary much, meaning the model consistently ranks positive samples higher than negatives. 

% Fourth paragraph (may be two paragraphs): Finally, reference Table 2 directly and discuss the trends. Which methods are best? How do they compare against the baselines? What are the insights that one should take away from Table 2? Describe in detail how you ran your significance test. Did you measure the mean of the metrics across local datasets or did you use k-fold cross validation to compute a mean? These things should be extremely clear to the point that someone could re-code your approach just from reading the paper. Also, if you're going to discuss non-iid and iid, make sure you define them in this context either here or in the methods before discussing them). 

Inspecting the results in Table~\ref{tab:res_baseline}, \textit{FedAvg} exhibits a slightly worse performance than the \textit{centralized} baseline, unveiling the fact that SLNs data are non-IID and training a single global model and applying it across all the SLNs, as done in vanilla \textit{FedAvg}, may not be an optimal choice. Nevertheless, the results demonstrate that as we move toward the pFL methods, we obtain performance gains over the \textit{centralized} baseline. These gains stem from the fact that pFL methods tailor individual models to the SLNs that can well-capture the local data characteristics and student interaction patterns in each SLN. In particular, \textit{FedALA} achieves the best performance across all the methods for all the classrooms/SLNs with the highest accuracy, lowest loss, and highest AUC.

% shows a gradual increase in model performance. Having trained on multiple classroom data with shared characteristics, \textit{FedAvg} is already capable of recognizing these shared patterns more than the \textit{Centralized} model, which can overfit to a single classroom based on data quantity. 
% By further finetuning the federated model to the class-specific patterns, \textit{FedAvg+FT} shows improved results compared to the FedAvg and centralized models for all datasets. %Finetuning further adapts the global model to the local characteristics of each class, leading to more accurate models. 
% However, with potential differences in student behavior, collaboration patterns, demographics, course dynamics, student-instructor relationships, or instructional styles across these classrooms contributing to their data being non-IID, implementing pFL offered better generalization. As a result, with better adaptation to local inherent characteristics, \textit{FedALA} shows the best accuracy across all classrooms with \textit{PerFedAvg-HF} generating competitive results. FedALA also has the lowest loss, indicating better convergence. AUC here is also the highest across classrooms, showing superior classification confidence. 
%For \textit{shake} and \textit{ml}, centralized and federated models perform almost the same, indicating that these datasets may be less affected by heterogeneity. 
% Therefore, can refine the details inherent in separate classes
\vspace{-1.5mm}
Assessing the \textit{centralized} model performance across various SLNs, \textit{csc} possesses the worst accuracy, indicating that its dataset highly differs --- both in terms of its size and its data patterns --- from other datasets. This suggests that \textit{csc} may benefit the most from the pFL methods.
The results confirm this hypothesis with \textit{csc} obtaining the highest performance improvements when comparing its accuracy between the \textit{centralized} and \textit{FedALA} methods --- approximate 5.54\% increase in accuracy and 19\% increase in AUC. Also, examining the overall results across all datasets reveals a consistent trend of improved performance when using pFL methods over the centralized baseline. In particular, the best-performing pFL method (i.e., FedALA) achieves an average improvement of 1.6\% in accuracy and a 4.2\% increase in mean AUC compared to the centralized model, highlighting the effectiveness of FL model personalization in handling diverse SLNs.

% 
% We notice at least a 19\% increase in AUC from baseline to FedALA model results for \textit{csc}, where loss values also decreased by at least 21\%.

% These improvements for the smallest dataset also indicate the impact of better local adaptation of class-specific characteristics to enhance model performance for the dataset, which had a weak influence on the baseline model due to its smaller size. 

% However, with the rest of the datasets dominating FedAvg updates, pFL did not improve their performances much from the FedAvg+FT results \cite{tamirisa2023fedselect}. 

%Nonetheless, our approach ensured a 1.6\% mean accuracy improvement over all classrooms with an average of 5.4\% decrease in loss and a 4.2\% mean AUC increase from the centralized model to FedALA. 

Conducting a paired t-test to assess the statistical significance of the performance differences with and without model personalization, we obtained a p-value less than 0.05 at a 95\% confidence interval. This result indicates that the improvements achieved by \textit{FedALA} over the \textit{centralized} model are statistically significant. Beyond this notable performance gain, \textit{FedALA} and all the other pFL methods offer the added benefit of preserving data privacy, further underscoring their practical advantages in real-world scenarios.

% In summary, our results highlight that pFL models not only improve performance but also fosters trust and encourages broader participation, making FL a superior choice for educational institutions. 

%Finetuning further adapts the global model to the local characteristics of each class, leading to more accurate models. 
%after finetuning with the added benefit of maintaining data privacy makes FL a superior choice for educational institutions  \cite{mcmahan2017communication, qin2023privacy}.   
%We chose a large batch size to process more samples in parallel, making training faster per epoch. 
%Figure~\ref{fig:curve} shows the learning curves of 
%We performed a grid search on hyperparameter values to find the best model for FL. %We saw more stable learning and the potential of not getting stuck at local minima with higher batch sizes and a higher number of SGD steps per round. 
 %A small batch size introduced gradient noise, which can prevent overfitting by better generalization.
 %All models reached convergence after training for [x] amount of time on average, where the system's computational power was [y]. 

\subsection{Fairness Evaluation}\label{sec:4-b}
% Results in Sec.~\ref{sec:4-a} show a clear indication that FL models outperform the baseline due to their ability to reduce noise from centralized aggregation, which is introduced by combining diverse datasets. %In centralized training, pooling data from different sources can distort class characteristics, leading to less accurate models. 
% However, to fully utilize the benefits of FL in education, it is imperative to understand how consistently and equitably FL performs across classrooms in comparison to other models discussed in this paper. In this section, we present fairness evaluation via \textit{equalized odds} or its more practical adaptation, \textit{equal opportunity}~\cite{hardt2016equality}, as described in Sec.~\ref{sec:3-e}.  

\begin{table}
\centering
\caption{TPR and FPR values for centralized, federated, and personalized FedALA models across different classrooms.}
{\small
\begin{tabular}{|c||c|c|c|}
\hline
\multirow{2}{*}{\textbf{Course}} & %\cellcolor{blue!60}
\textbf{Centralized} & %\cellcolor{red!60}
\textbf{FedAvg} & %\cellcolor{green!60}
\textbf{FedALA}\\
 & %\cellcolor{blue!60}
 \textbf{(TPR, FPR)} & %\cellcolor{red!60}
 \textbf{(TPR, FPR)}& %\cellcolor{green!60}
 \textbf{(TPR, FPR)}\\
\hline \hline %\hline
\cellcolor{green!20}algo & (0.83, 0.03) & (0.81, 0.05) & (0.84, 0.03)\\
\hline
\cellcolor{red!20}comp & (0.72, 0.04) & (0.76, 0.05) & (0.74, 0.04) \\
\hline
\cellcolor{blue!20}shake & (0.72, 0.05) & (0.77, 0.07) & (0.68, 0.04) \\
\hline
\cellcolor{orange!20}ml & (0.81, 0.1) & (0.71, 0.08) & (0.85, 0.11) \\
\hline
\cellcolor{gray!20}csc & (0.28, 0.02) & (0.37, 0.03) & (0.68, 0.05) \\
\hline
\end{tabular}
}
\label{tab:fairness_comparison}
\end{table}

Next, we examine the fairness of the models obtained from different methods, based on the methodology outlined in Sec.~\ref{sec:3-e}. 
To this end, we compare how close the TPR and FPR were across the classrooms/SLNs through the values presented in Table~\ref{tab:fairness_comparison}.
In this comparison, we focus solely on the best-performing pFL model identified in Table~\ref{tab:res_baseline}, namely \textit{FedALA}, to highlight how personalization impacts fairness relative to the vanilla \textit{FedAvg} and \textit{centralized} methods.

% where we have only kept the best performing pFL model from the results obtained in Table~\ref{tab:res_baseline} (i.e., FedALA) in the comparisons. 

\vspace{-1mm}
\begin{figure}[t]
    \centering
    \begin{subfigure}[b]{0.23\textwidth}  % Adjust width slightly
        \centering
        \includegraphics[width=.9\linewidth, alt={The range (difference between the highest and lowest) of TPR values across all classrooms/SLNs.}]{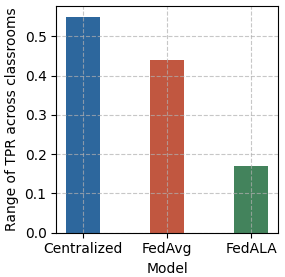}
        \caption{TPR range}
        \label{fig:fairness tpr}
    \end{subfigure}
    % \hfill  % Ensures spacing between subfigures
    \begin{subfigure}[b]{0.23\textwidth}  
        \centering
        \includegraphics[width=.9\linewidth, alt={The range (difference between the highest and lowest) of FPR values across all classrooms/SLNs.}]{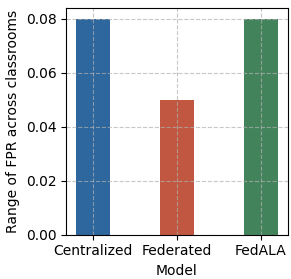}
        \caption{FPR range}
        \label{fig:fairness fpr}
    \end{subfigure}
    \caption{The range (difference between the highest and lowest) of TPR and FPR values across all classrooms/SLNs.}
     \label{fig:fairness}
        %Comparison of TPR values across classrooms shows the least range (\textit{max-min}) for FedALA (left sub-plot), while its FPR value range is on par with that of the centralized model.}
\end{figure}

% \begin{figure}[ht]
%     \centering   
%     \includegraphics[width=0.9\linewidth, alt={TPR/FPR range comparison across classrooms}]{fair.png}
%         \caption{The range (difference between the highest and lowest) of TPR values (left sub-plot) and FPR values (right sub-plot) across all classrooms/SLNs.}%Comparison of TPR values across classrooms shows the least range (\textit{max-min}) for FedALA (left sub-plot), while its FPR value range is on par with that of the centralized model.}
%         \label{fig:fairness}
% \end{figure}
\vspace{-1mm}
% \begin{figure}
%     \centering
%     \includegraphics[width=0.45\linewidth, alt={TPR range comparison across classrooms}]{tpr.png}
%     \caption{Equal opportunity difference metric shows the least range (difference between the highest and lowest TPR across all classrooms/SLNs) for FedALA.}
%     \label{fig:fairness}
% \end{figure}

We first focus on the TPR values in Table~\ref{tab:fairness_comparison} and observe a higher consistency across the TPR values for \textit{FedALA} compared to both \textit{centralized} and \textit{FedAvg} methods. This implies that \textit{FedALA} adheres to the equal opportunity measure more closely. To quantify this finding, we compute the equal opportunity difference (EOD) metric \cite{ezzeldin2023fairfed}, which measures the difference between TPR values across the classrooms, referred to as the \textit{range} of TPR. A smaller \textit{range} of TPR is desirable as it reflects more consistent positive predictions across different student pairs belonging to different SLNs, contributing to model fairness --- the most fair model would have the \textit{range} of $0$. We depict the \textit{range} of TPR of the methods in Figure \ref{fig:fairness}(a), which verifies that \textit{FedALA} achieves the lower TPR \textit{range}, implying its higher adherence to equal opportunity measure.  
On the other hand, inspecting the FPR values in Table~\ref{tab:fairness_comparison} there seem to be no conclusive results: the range (the difference between the maximum and minimum value computed over the SLNs) of FPR for \textit{FedALA} is on par with \textit{centralized} but is worse than \textit{FedAvg} as also observed in Figure \ref{fig:fairness}(b). So, a concrete conclusion on the equal odds measure cannot be drawn from the results. In summary, by exhibiting a better TPR consistency performance in terms of EOD, \textit{FedALA} outperforms the other methods in terms of model fairness. Also, the results open a door to future studies on improving the adherence of FL models to the equal odds measure in SLNs.

\vspace{-1mm}
\subsection{Contributions of Features in Predictions}\label{sec:4-c}
We next discuss the explainability of the models, based on the methodology outlined in Sec.~\ref{sec:3-f}.
% In this section, we offer explainability of the models discussed, as outlined in Sec.~\ref{sec:3-f}. 
We showcase Shapley values in different SHAP plots that illustrate the importance/impact of each feature on the models' predictions. In the following, we study the SHAP plots in terms of both individual level interpretations for a student pair as well as global level interpretations for all student pairs.

% \begin{figure*}[ht]
%     \centering
%     \begin{subfigure}[b]{\textwidth}  % Adjust width slightly
%         \centering
%         \includegraphics[width=\linewidth]{force_algo_cen.png}
%         \caption{Centralized}
%         \label{fig:algo-cen-force}
%     \end{subfigure}
%     % \hfill  % Ensures spacing between subfigures
%     \begin{subfigure}[b]{\textwidth}  
%         \centering
%         \includegraphics[width=\linewidth]{force_algo_ft.png}
%         \caption{Finetuned}
%         \label{fig:algo-ft-force}
%     \end{subfigure}
%     \caption{Force plots for an individual learner pair belonging to \textit{algo} classroom present the relationship between feature values and prediction based on two models as shown in (a) and (b).}
% \end{figure*}

\begin{figure*}[!h]
    \centering
    \includegraphics[width=\linewidth, alt={Force plot for a student in algo}]{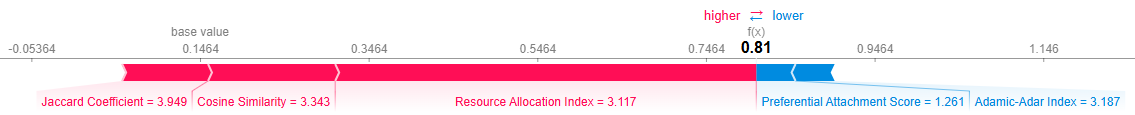}
    \caption{Force plot for a learner pair in \textit{algo} classroom, presenting the relationship between feature values and prediction.}% by the finetuned model.}
    \label{fig:algo-ft-force}
\end{figure*}

\begin{figure*}[ht]
    \centering
    \begin{subfigure}[b]{0.48\textwidth}  % Adjust width slightly
        \centering
        \includegraphics[width=\linewidth, alt={Feature importance plot for algo}]{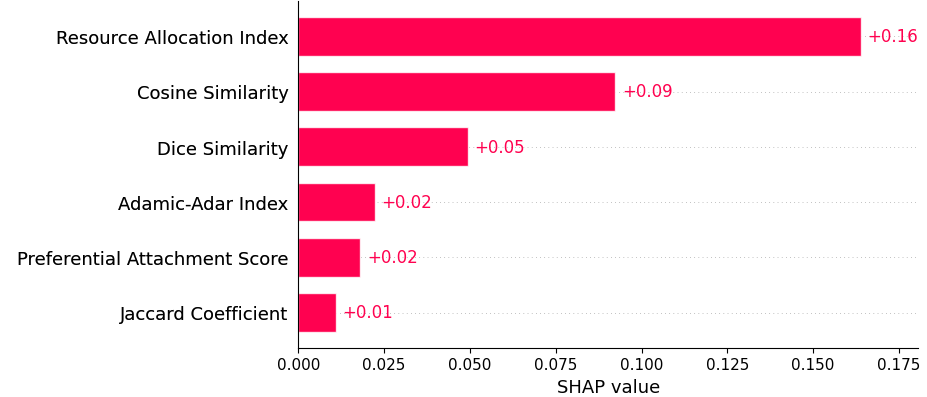}
        \caption{algo}
        \label{fig:algo-shap}
    \end{subfigure}
    \hfill  % Ensures spacing between subfigures
    \begin{subfigure}[b]{0.48\textwidth}  
        \centering
        \includegraphics[width=\linewidth, alt={Feature importance plot for ml}]{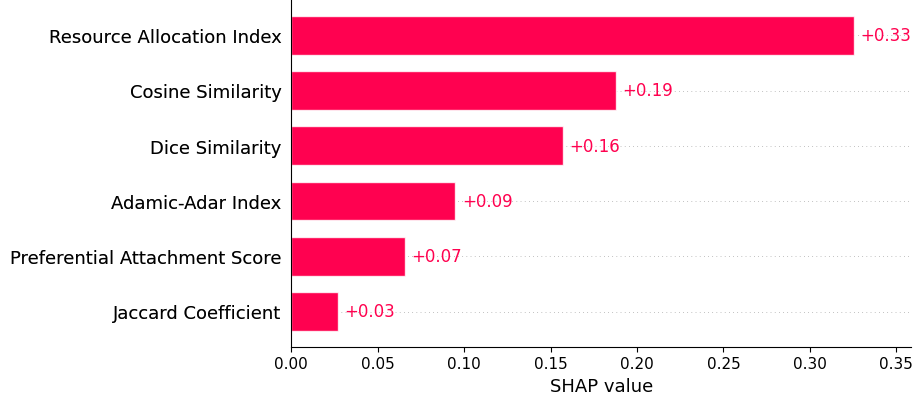}
        \caption{ml}
        \label{fig:ml-shap}
    \end{subfigure}
    \hfill
    \begin{subfigure}[b]{0.48\textwidth}
        \centering
        \includegraphics[width=\linewidth, alt={Feature importance plot for comp}]{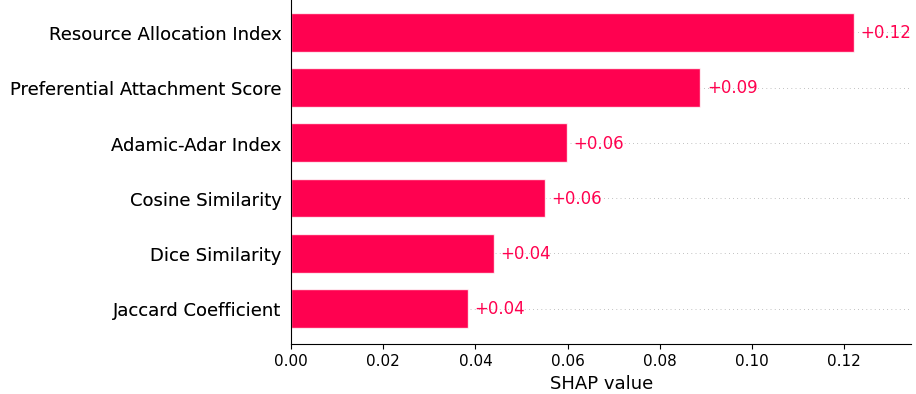}
        \caption{comp}
        \label{fig:comp-shap}
    \end{subfigure}
    \hfill
    \begin{subfigure}[b]{0.48\textwidth}
        \centering
        \includegraphics[width=\linewidth, alt={Feature importance plot for shake}]{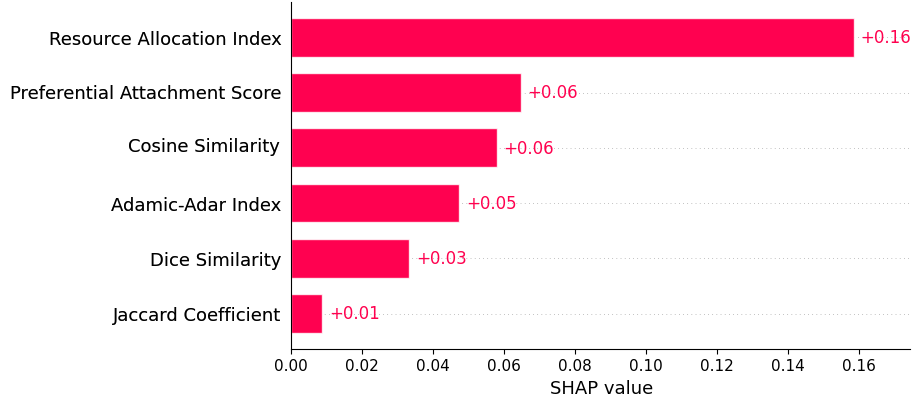}
        \caption{shake}
        \label{fig:shake-shap}
    \end{subfigure}
    \caption{Relative importance of features across all MOOC SLNs show the same set of top-2 features for (a) and (b), which are STEM courses, and the same set of top-2 features for (c) and (d), which are non-STEM courses.}
    \label{fig:imp}
\end{figure*}
\begin{figure}[ht]
    \centering
    \includegraphics[width=\linewidth, alt={Feature importance plot for csc}]{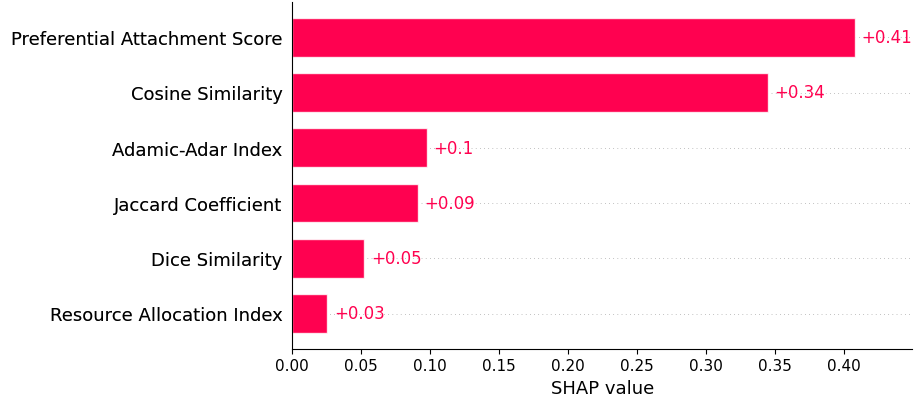}
    \caption{Feature importance in STEM \& non-MOOC \textit{csc}.}
    \label{fig:csc-shap}
\end{figure}
\vspace{-3mm}
\begin{figure*}[ht]
    \centering
    \begin{subfigure}[b]{0.48\textwidth}  % Adjust width slightly
        \centering
        \includegraphics[width=\linewidth, alt={Summary plot for algo with centralized model}]{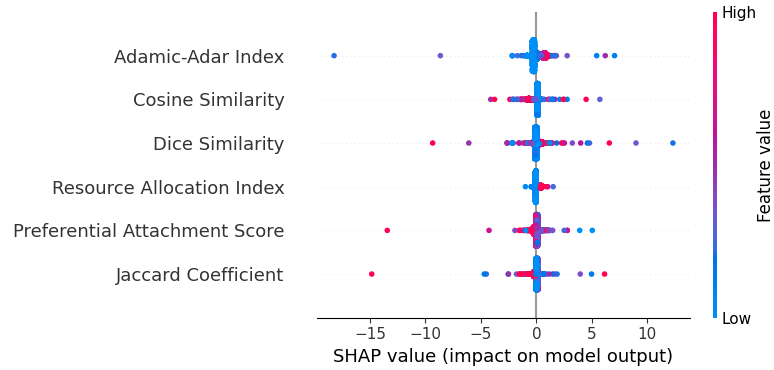}
        \caption{Centralized}
        \label{fig:algo-cen}
    \end{subfigure}
    \hfill  % Ensures spacing between subfigures
    \begin{subfigure}[b]{0.48\textwidth}  
        \centering
        \includegraphics[width=\linewidth, alt={Summary plot for algo with fine-tuned model}]{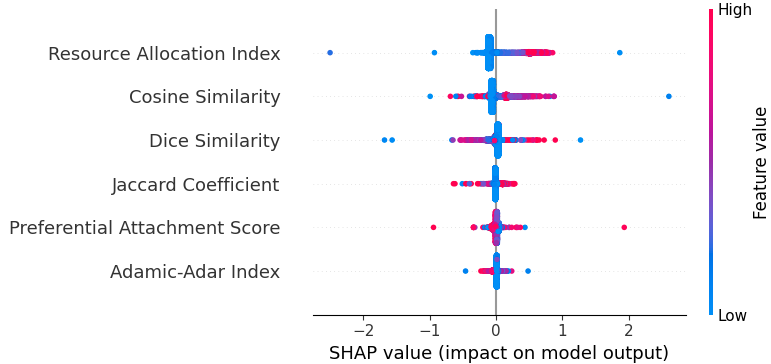}
        \caption{FedALA}
        \label{fig:algo-ft}
    \end{subfigure}
    \caption{Feature importance explainability plots for the \textit{algo} classroom under (a) centralized and (b) FedALA model training. The results unveil a change in the significance/contribution of individual features from the \textit{centralized} to \textit{FedALA} model.}
        \label{fig:summary}
\end{figure*}
% \vspace{-1cm}
% \vspace{-1mm}
\subsubsection{Individual Level Explanation}
SHAP allows us to perform data instance-specific explanations based on the computed Shapley values, where each feature gets a score representing its contribution to moving the prediction away from the base value (mean prediction value for the dataset).
Focusing on the predictions made by \textit{FedALA} model, 
Figure \ref{fig:algo-ft-force} shows the most important features and their corresponding values for a linked student pair in \textit{algo} class. The base value (i.e., 0.1464) represents the average model output across the dataset before incorporating individual feature effects. The final predicted value is 0.81, which is much higher than the base value, meaning that the contributing features collectively increased the prediction. Each feature's impact is visualized by arrows: the longer the arrow for a feature, the greater the impact of that feature on the prediction. 
Note that the red-colored features increased the link formation probability higher, while the blue-colored features did the opposite. As can be seen, high \textit{Jaccard Coefficient, Cosine Similarity}, and \textit{Resource Allocation Index} values helped the model in its predictions, whereas low \textit{Preferential Attachment Score} and \textit{Adamic-Adar Index} values have often lowered the prediction. 

This observation highlights two important points: (i) features \textit{do not} contribute equally to the model’s predictions, and (ii) there may be \textit{significant differences} in the contribution of individual features, underscoring the importance of understanding feature relevance for informed model interpretation. This motivates our subsequent discussions, where we discuss the importance of the features between the pairs of students in all classrooms and investigate their impacts.

% These observations indicate that with a high proportion of shared neighbors, these two students are highly likely to connect. However, not all highly connected students as common neighbors necessarily promote link formation. Finally, with the final prediction pushed up to 0.81, it is evident that the model saw a high probability of link formation between this pair. Intuitively, this phenomenon in this student pair means that the more connections made individually, the more chances that they will pair up. In the following section, we discuss the importance of the features between the pairs of students in all classrooms and investigate their impact.
\vspace{-2.3mm}
\subsubsection{Global Level Explanation}
% With an initial hypothesis that non-STEM classrooms tend to have less defined interaction structures, with students often responding to the most engaged peers; whereas, in contrast, STEM classrooms show stronger and more stable connections between students who interact regularly, 

We next provide dataset-wide explanations on trends of feature importance, focusing on the FedALA model. In Figures~\ref{fig:imp}(a)-(d), and \ref{fig:csc-shap}, we illustrate the relative feature importance using SHAP plots to assess feature contributions at a global level for all student pairs for both MOOC and non-MOOC courses/SLNs.
The X-axis represents each feature's relative impact on the model's predictions. We summarize the main takeaways of these results below:

% \ali{HERE! BRB!}

% \begin{description}
\vspace{-1mm}
    \textbf{STEM and MOOC SLNs (\textit{algo} \& \textit{ml}):} \textit{Resource Allocation Index} and \textit{Cosine Similarity} (Figures \ref{fig:imp}(a) \& (b)) have the most impact on the model prediction performance. 
    Inspecting the logic behind these two metrics, discussed in~Sec.~\ref{sec:feature}, this outcome matches the intuition that STEM classrooms expect students to collaborate for learning and achieving success, and thus any two students working on similar assignments will likely share several mutual peers, increasing the probability of their direct collaboration. Furthermore, two students with similar coding performance on assignments will likely discuss solutions in MOOC forums or work together on projects. On the other hand, in a MOOC format course, due to the lack of face-to-face interactions, students tend to have more diverse conversations with random peers. Hence, two students who communicate with similar neighbors more exclusively have a higher potential to connect as part of the same online sub-community.
\vspace{-1mm}

    \textbf{Non-STEM and MOOC SLNs (\textit{comp} \& \textit{shake}):} \textit{Resource Allocation Index} and \textit{Preferential Attachment Score} (Figures \ref{fig:imp}(c) \& (d)) have the most impact on the model performance. Inspecting the logic behind these two metrics, discussed in~Sec.~\ref{sec:feature}, this outcome also matches the intuition that non-STEM classrooms encourage open-ended discussions, and thus, students might frequently respond to different peers in a MOOC discussion forum. As a result, a highly active pair of students has a higher probability of corresponding with one another. Also, similar to the STEM and MOOC SLNs discussed above, since correspondences in the MOOC format are more diverse, a high \textit{Resource Allocation Index} can demonstrate students who have formed an online community.
\vspace{-2.2mm}    

    \textbf{STEM and non-MOOC SLN (\textit{csc}):}~\textit{Preferential Attachment Score} and \textit{Cosine Similarity} (Figure \ref{fig:csc-shap}) have the most impact on the model performance. Inspecting the logic behind these two metrics, discussed in~Sec.~\ref{sec:feature}, this outcome matches the intuition that in a non-MOOC STEM classroom, students tend to work in structured groups formed during in-person interactions, and thus, the proportion of shared neighbors plays an important role in increasing the possibility of future link formation. Furthermore, since online correspondence in this type of class is less compared to non-STEM and MOOC, students who have a tendency to be active online have a higher tendency to communicate.
    
    % In a classroom-based CS course, any two students with similar performance in coding assignments are likely to be assigned to the same homework team. In addition, students who frequently work with many others tend to receive more connections over time.     
% \end{description}

In summary, \textit{Resource Allocation Index} turns out to be the most prominent feature for MOOC courses --- implying that common neighbors and neighborhood influence forming connections in these settings. Interestingly, this feature is the least important for the non-MOOC course --- implying less reliance on shared connections to predict interactions. Such differences in feature importance can highlight classrooms' unique needs and aid instructors with intervention guidance.%, e.g., customized learning resources or collaborative assignments.% that can aid in policy intervention for instructors and institutions. 

To further understand the relationship between feature values and predicted outcomes and how it may differ upon conducting model personalization, we present summary plots for all student pairs in \textit{algo} classroom in Figure~\ref{fig:summary}. Each point in the plot represents the Shapley value for a student pair, and features are sorted according to their importance. Data points with overlapping Shapley values oscillate along the Y-axis. The color gradient reflects feature values, with red indicating high values and blue representing low values. Between the \textit{centralized} and \textit{FedALA} models, we observe a notable change in the order of the features. This reveals a nuanced phenomenon: \textit{FedALA obtains the performance gains reported in Table~\ref{tab:res_baseline} by putting a different emphasis on the features, which are tailored to each SLN.} Interestingly, from being the most important feature for the \textit{centralized} model, \textit{Adamic-Adar Index} becomes the least important feature after personalization in \textit{FedALA}, implying the shift of focus to \textit{Resource Allocation Index} that boosts low-degree nodes more aggressively than \textit{Adamic-Adar Index} for this MOOC STEM course. These findings can be intuitively explained further in the context of an algorithm online course, where learners do not randomly connect with others but build their network based on shared connections. This might happen as learners trust and engage with those connected to their existing network. As a result, students with common neighbors are more likely to connect in discussions, projects, and coding forums, leading to better engagement and learning success. These observations can offer distinguishing patterns that have the potential to help instructors take more control and provide better learning environments.   

% The results on other datasets, which are omitted due to them being qualitatively similar to Figure~\ref{fig:summary}, similarly demonstrated that \textit{FedALA} model captures the unique interaction dynamics or features that are more relevant within the classrooms.
The results from other datasets, which are omitted here due to their qualitative similarity to Figure~\ref{fig:summary}, similarly demonstrate that the FedALA model effectively captures the unique interaction dynamics and identifies features that are more relevant within each classroom. 
\vspace{-1mm}
\subsection{Implications for Design}\label{sec:4-d}
The experimental results collectively showed that FL can learn distinct behavioral patterns for students in each classroom, especially when fine-tuned/personalized locally for better adaptation to the individual data patterns in each SLN. This alleviates the need for using a single centralized model that needs direct access to data from all classrooms, which can cause privacy concerns.
The results further provide opportunities to improve students' learning experiences through timely and effective interventions.
For instance, link prediction can help identify students who may engage less socially, allowing teachers to target interventions like pairing them with peers to foster collaboration. Such interventions can help struggling students, promote stronger peer connections, and increase overall class performance \cite{berlinski2023helping, li2024math}. Another example is when students who may benefit the most from collaborative study or project groups are identified. For example, the odds of students forming links can be correlated with their personal characteristics, such as academic strengths, learning styles, engagement patterns, etc. Coupled with their class performance measures, the results can further inspire methods to identify characteristics of successful collaboration among students, leading to intentional pairings that can lead to more balanced and effective group dynamics.

% While there is much room to research the best grouping approaches, social network analysis in classrooms can help identify characteristics that facilitate effective collaboration among students. For example, the odds of students forming links can be correlated with their personal characteristics, such as academic strengths, learning styles, engagement patterns, etc. Coupled with their class performance measures, we can identify characteristics of successful collaboration. Such intentional pairing can lead to more balanced and effective group dynamics, with a stronger peer-to-peer support environment. 
\vspace{-2mm}
\section{Limitations \& Future Work}

Focusing on the limitation of our work, the datasets used for this study did not have student demographic information due to data anonymization practices,  preventing us from fine-tuning and testing our pFL models in various student demographic groups. We aim to work on this in the future to demonstrate FL's potential to address the bias against underrepresented/minority students. 

Focusing on future work, investigation of the relationship between predicted links and actual student learning gains in a real-world scenario is a promising research direction. In addition, it is worth studying few-shot learning techniques~\cite{guo2017one, aguirre2023selecting} to adapt models trained in large classrooms in the pFL setting to smaller or underrepresented ones with limited data. Furthermore, using large language models (LLMs) \cite{zhao2023survey, xu2024large, yang2023palr, zytek2024explingo} to (a) generate human-readable explanations of link predictions, aiding educators in understanding student networks, (b) extract rich semantic features from students' forum posts, chat logs, and written submissions for improved link prediction, and (c) suggest personalized interventions (e.g., peer group formations) based on student interaction histories, all are exciting research opportunities.

% Third, data heterogeneity might have posed a challenge in achieving fairness, which is an inherent limitation of FL environments operating on decentralized and non-IID data. 
% However, the goal of our paper is not to resolve these broader issues but to contribute to the understanding of FL under real-world conditions where data distributions across clients can vary, and we want to protect the privacy of each entity in an educational setting that is overlooked in a centralized model. Addressing fairness in these settings will be a key focus of our future research, as our findings highlight the need for further exploration in this area. %, as well as more complex methods to model the underlying patterns \cite{potgieter2008temporality}. %Mining data from consecutive semesters from educational institutions can solve this shortcoming. 
% Furthermore, we will explore more advanced personalized FL approaches (FedRep \cite{collins2021exploiting}, MetaFed \cite{chen2023metafed}, 
% and Fedl2p \cite{lee2024fedl2p}, among others) next to offer even better adaptation, increasing accuracy and fairness in heterogeneous classroom settings. 
\vspace{-2mm}
\section{Conclusion}

This study presented a novel application of federated learning (FL) for link prediction in social learning networks (SLNs). By modeling SLNs from multiple classrooms as graph structures representing student interactions, we computed topological features for student pairs and utilized them for model training and testing.
Unlike centralized training methods that require aggregating raw data, our FL-based approach allowed each classroom/SLN to train its model locally while only transmitting model parameters to a server for aggregation, preserving the privacy of student interaction data.
To further enhance performance, we incorporated model personalization techniques, enabling the FL model to adapt to the distinctive characteristics of individual classrooms. Our experimental results demonstrated that personalized FL models outperformed both vanilla FL and centralized models, highlighting the impact of unique student interaction patterns inherent to individual SLNs.
In addition, we explored the explainability of the trained models using explainable AI (XAI) techniques. This analysis revealed differences in the importance of various topological features for link prediction across different classroom types (e.g., STEM vs. non-STEM, MOOC vs. non-MOOC), offering valuable insights into how interaction patterns vary by context. 

\bibliographystyle{abbrv}
\bibliography{sigproc}  % sigproc.bib is the name of the Bibliography in this case

\begin{thebibliography}{10}

\bibitem{FERPA}
Family educational rights and privacy act ({FERPA}).
\newblock https://epic.org/family-educational-rights-and-privacy-act-ferpa/, Online; accessed 19 Feb. 2025.

\bibitem{adamic2003friends}
L.~A. Adamic and E.~Adar.
\newblock Friends and neighbors on the web.
\newblock {\em Social networks}, 25(3):211--230, 2003.

\bibitem{aghababaei2019interpolative}
S.~Aghababaei and M.~Makrehchi.
\newblock Interpolative self-training approach for link prediction.
\newblock {\em Intelligent Data Analysis}, 23(6):1379--1395, 2019.

\bibitem{aguirre2023selecting}
C.~Aguirre, K.~Sasse, I.~Cachola, and M.~Dredze.
\newblock Selecting shots for demographic fairness in few-shot learning with large language models.
\newblock {\em arXiv preprint arXiv:2311.08472}, 2023.

\bibitem{amershi2009combining}
S.~Amershi, C.~Conati, et~al.
\newblock Combining unsupervised and supervised classification to build user models for exploratory learning environments.
\newblock {\em Journal of educational data mining}, 1(1):18--71, 2009.

\bibitem{ammad2019federated}
M.~Ammad-Ud-Din, E.~Ivannikova, S.~A. Khan, W.~Oyomno, Q.~Fu, K.~E. Tan, and A.~Flanagan.
\newblock Federated collaborative filtering for privacy-preserving personalized recommendation system.
\newblock {\em arXiv preprint arXiv:1901.09888}, 2019.

\bibitem{asad2021federated}
M.~Asad, A.~Moustafa, and T.~Ito.
\newblock Federated learning versus classical machine learning: A convergence comparison.
\newblock {\em arXiv preprint arXiv:2107.10976}, 2021.

\bibitem{backstrom2011supervised}
L.~Backstrom and J.~Leskovec.
\newblock Supervised random walks: predicting and recommending links in social networks.
\newblock In {\em Proceedings of the fourth ACM international conference on Web search and data mining}, pages 635--644, 2011.

\bibitem{baek2023personalized}
J.~Baek, W.~Jeong, J.~Jin, J.~Yoon, and S.~J. Hwang.
\newblock Personalized subgraph federated learning.
\newblock In {\em International conference on machine learning}, pages 1396--1415. PMLR, 2023.

\bibitem{bandura1977social}
A.~Bandura.
\newblock Social learning theory.
\newblock {\em Englewood Cliffs}, 1977.

\bibitem{barabasi1999emergence}
A.-L. Barab{\'a}si and R.~Albert.
\newblock Emergence of scaling in random networks.
\newblock {\em science}, 286(5439):509--512, 1999.

\bibitem{bashiri2024transformative}
M.~Bashiri and K.~Kowsari.
\newblock Transformative influence of llm and ai tools in student social media engagement: Analyzing personalization, communication efficiency, and collaborative learning.
\newblock {\em arXiv preprint arXiv:2407.15012}, 2024.

\bibitem{berlinski2023helping}
S.~Berlinski, M.~Busso, and M.~Giannola.
\newblock Helping struggling students and benefiting all: Peer effects in primary education.
\newblock {\em Journal of Public Economics}, 224:104925--104948, 2023.

\bibitem{bhattacharya2023towards}
S.~Bhattacharya, P.~S. Vyas, S.~Yarradoddi, B.~Dasari, S.~Sumathy, R.~Kaluri, S.~Koppu, D.~J. Brown, M.~Mahmud, and T.~R. Gadekallu.
\newblock Towards smart education in the industry 5.0 era: A mini review on the application of federated learning.
\newblock In {\em Proceedings of the 2023 IEEE International Conference on Dependable, Autonomic and Secure Computing, International Conference on Pervasive Intelligence and Computing, International Conference on Cloud and Big Data Computing, International Conference on Cyber Science and Technology Congress (DASC/PiCom/CBDCom/CyberSciTech)}, pages 602--608. IEEE, 2023.

\bibitem{bojanowski2020proximity}
M.~Bojanowski and B.~Chro{\l}.
\newblock Proximity-based methods for link prediction in graphs with r package'linkprediction'.
\newblock {\em ASK. Research and Methods}, 29(1):5--28, 2020.

\bibitem{brinton2018efficiency}
C.~G. Brinton, S.~Buccapatnam, L.~Zheng, D.~Cao, A.~S. Lan, F.~M. Wong, S.~Ha, M.~Chiang, and H.~V. Poor.
\newblock On the efficiency of online social learning networks.
\newblock {\em IEEE/ACM Transactions on Networking}, 26(5):2076--2089, 2018.

\bibitem{brown2015good}
R.~Brown, C.~F. Lynch, M.~Eagle, J.~L. Albert, T.~Barnes, R.~S. Baker, Y.~Bergner, and D.~S. McNamara.
\newblock Good communities and bad communities: Does membership affect performance?
\newblock In O.~C. Santos, J.~Boticario, C.~Romero, M.~Pechenizkiy, A.~Merceron, P.~Mitros, J.~M. Luna, M.~C. Mihaescu, P.~Moreno, A.~Hershkovitz, S.~Ventura, and M.~C. Desmarais, editors, {\em Proceedings of the 8th International Conference on Educational Data Mining, {EDM} 2015, Madrid, Spain, June 26-29, 2015}, pages 612--613. International Educational Data Mining Society {(IEDMS)}, 2015.

\bibitem{carolan2013social}
B.~V. Carolan.
\newblock {\em Social network analysis and education: Theory, methods \& applications}.
\newblock Sage Publications, 2013.

\bibitem{choudhary2023social}
S.~Choudhary, K.~Sharma, and M.~Bajaj.
\newblock Social networks analysis and machine learning: an overview of approaches and applications.
\newblock In {\em 2023 International Conference on Sustainable Computing and Smart Systems (ICSCSS)}, pages 123--128, 2023.

\bibitem{chu2022mitigating}
Y.-W. Chu, S.~Hosseinalipour, E.~Tenorio, L.~Cruz, K.~Douglas, A.~Lan, and C.~Brinton.
\newblock Mitigating biases in student performance prediction via attention-based personalized federated learning.
\newblock In {\em Proceedings of the 31st ACM International Conference on Information \& Knowledge Management}, pages 3033--3042, 2022.

\bibitem{chu2024multi}
Y.-W. Chu, S.~Hosseinalipour, E.~Tenorio, L.~Cruz, K.~Douglas, A.~S. Lan, and C.~G. Brinton.
\newblock Multi-layer personalized federated learning for mitigating biases in student predictive analytics.
\newblock {\em IEEE Transactions on Emerging Topics in Computing}, X:1--15, 2024.

\bibitem{tungar2023evaluation}
T.~D.~V. and P.~D.~V.
\newblock Evaluation of privacy-preserving techniques: Bouncy castle encryption and machine learning algorithms for secure classification of sensitive data.
\newblock {\em International Journal of Intelligent Systems and Applications in Engineering}, 11:429 –, Jul. 2023.

\bibitem{dice1945measures}
L.~R. Dice.
\newblock Measures of the amount of ecologic association between species.
\newblock {\em Ecology}, 26(3):297--302, 1945.

\bibitem{ebrahimi2025transition}
M.~Ebrahimi, R.~Sahay, S.~Hosseinalipour, and B.~Akram.
\newblock The transition from centralized machine learning to federated learning for mental health in education: A survey of current methods and future directions.
\newblock {\em arXiv preprint arXiv:2501.11714}, 2025.

\bibitem{ezzeldin2023fairfed}
Y.~H. Ezzeldin, S.~Yan, C.~He, E.~Ferrara, and A.~S. Avestimehr.
\newblock Fairfed: Enabling group fairness in federated learning.
\newblock In {\em Proceedings of the AAAI conference on artificial intelligence}, volume~37, pages 7494--7502, 2023.

\bibitem{fallah2020personalized}
A.~Fallah, A.~Mokhtari, and A.~Ozdaglar.
\newblock Personalized federated learning with theoretical guarantees: A model-agnostic meta-learning approach.
\newblock {\em Advances in neural information processing systems}, 33:3557--3568, 2020.

\bibitem{gharahighehi2022addressing}
A.~Gharahighehi, K.~Pliakos, and C.~Vens.
\newblock Addressing the cold-start problem in collaborative filtering through positive-unlabeled learning and multi-target prediction.
\newblock {\em IEEE Access}, 10:117189--117198, 2022.

\bibitem{guo2017one}
Y.~Guo and L.~Zhang.
\newblock One-shot face recognition by promoting underrepresented classes.
\newblock {\em arXiv preprint arXiv:1707.05574}, 2017.

\bibitem{gurjar2020leveraging}
N.~Gurjar.
\newblock Leveraging social networks for authentic learning in distance learning teacher education.
\newblock {\em TechTrends}, 64(4):666--677, 2020.

\bibitem{hardt2016equality}
M.~Hardt, E.~Price, and N.~Srebro.
\newblock Equality of opportunity in supervised learning.
\newblock {\em Advances in neural information processing systems}, 29:3323--3331, 2016.

\bibitem{hasan2011survey}
M.~A. Hasan and M.~J. Zaki.
\newblock {\em A Survey of Link Prediction in Social Networks}, pages 243--275.
\newblock Springer US, Boston, MA, 2011.

\bibitem{hollister2022engagement}
B.~Hollister, P.~Nair, S.~Hill-Lindsay, and L.~Chukoskie.
\newblock Engagement in online learning: student attitudes and behavior during covid-19.
\newblock In C.~Blake, editor, {\em Frontiers in education}, volume~7, page 851019, Lausanne, Switzerland, 2022. Frontiers Media SA.

\bibitem{hridi2024revolutionizing}
A.~P. Hridi, R.~Sahay, S.~Hosseinalipour, and B.~Akram.
\newblock Revolutionizing ai-assisted education with federated learning: A pathway to distributed, privacy-preserving, and debiased learning ecosystems.
\newblock In {\em Proceedings of the AAAI Symposium Series}, volume~3, pages 297--303, 2024.

\bibitem{jaccard1912distribution}
P.~Jaccard.
\newblock The distribution of the flora in the alpine zone. 1.
\newblock {\em New phytologist}, 11(2):37--50, 1912.

\bibitem{jiang2022privacy}
Z.~L. Jiang, J.~Gu, H.~Wang, Y.~Wu, J.~Fang, S.-M. Yiu, W.~Luo, and X.~Wang.
\newblock Privacy-preserving distributed machine learning made faster, 2022.

\bibitem{jie2019social}
Z.~Jie, Y.~Li, and R.~Liu.
\newblock Social network group identification based on local attribute community detection.
\newblock In {\em 2019 IEEE 3rd Information Technology, Networking, Electronic and Automation Control Conference (ITNEC)}, pages 443--447. IEEE, 2019.

\bibitem{kairouz2021advances}
P.~Kairouz, H.~B. McMahan, B.~Avent, A.~Bellet, M.~Bennis, A.~N. Bhagoji, K.~Bonawitz, Z.~Charles, G.~Cormode, R.~Cummings, et~al.
\newblock Advances and open problems in federated learning.
\newblock {\em Foundations and trends{\textregistered} in machine learning}, 14(1--2):1--210, 2021.

\bibitem{kanselaar2002constructivism}
G.~Kanselaar.
\newblock Constructivism and socio-constructivism.
\newblock {\em Journal of Educational Psychology}, 94(3):1--10, 2002.

\bibitem{khelghatdoust2022socially}
M.~Khelghatdoust and M.~Mahdavi.
\newblock A socially-aware, privacy-preserving, and scalable federated learning protocol for distributed online social networks.
\newblock In {\em International Conference on Advanced Information Networking and Applications}, pages 192--203. Springer, 2022.

\bibitem{li2024math}
H.~Li, S.~Zhang, S.~Lee, J.-E. Lee, Z.~Zhong, E.~Weitnauer, and A.~F. Botelho.
\newblock Math in motion: Analyzing real-time student collaboration in computer-supported learning environments.
\newblock In {\em Proceedings of the 17th International Conference on Educational Data Mining}, pages 533--541, 2024.

\bibitem{liben2003link}
D.~Liben-Nowell and J.~Kleinberg.
\newblock The link prediction problem for social networks.
\newblock In {\em Proceedings of the twelfth International Conference on Information and Knowledge Management}, pages 556--559, 2003.

\bibitem{liu2022distributed}
J.~Liu, J.~Huang, Y.~Zhou, X.~Li, S.~Ji, H.~Xiong, and D.~Dou.
\newblock From distributed machine learning to federated learning: A survey.
\newblock {\em Knowledge and Information Systems}, 64(4):885--917, 2022.

\bibitem{liu2022federated}
Z.~Liu, L.~Yang, Z.~Fan, H.~Peng, and P.~S. Yu.
\newblock Federated social recommendation with graph neural network.
\newblock {\em ACM Transactions on Intelligent Systems and Technology (TIST)}, 13(4):1--24, 2022.

\bibitem{lu2011link}
L.~L{\"u} and T.~Zhou.
\newblock Link prediction in complex networks: A survey.
\newblock {\em Physica A: statistical mechanics and its applications}, 390(6):1150--1170, 2011.

\bibitem{lundberg2017unified}
S.~Lundberg.
\newblock A unified approach to interpreting model predictions.
\newblock {\em arXiv preprint arXiv:1705.07874}, 2017.

\bibitem{ma2024mixture}
L.~Ma, H.~Han, J.~Li, H.~Shomer, H.~Liu, X.~Gao, and J.~Tang.
\newblock Mixture of link predictors on graphs.
\newblock {\em arXiv preprint arXiv:2402.08583}, 2024.

\bibitem{mcmahan2017communication}
H.~B. McMahan, E.~Moore, D.~Ramage, S.~Hampson, and B.~Aguera~y Arcas.
\newblock Communication-efficient learning of deep networks from decentralized data.
\newblock In A.~Singh and J.~Zhu, editors, {\em Proceedings of the 20th International Conference on Artificial Intelligence and Statistics (AISTATS)}, volume~54 of {\em Proceedings of Machine Learning Research}, pages 1273--1282, Fort Lauderdale, FL, USA, Apr. 2017. PMLR.

\bibitem{mezghani2018online}
M.~Mezghani, M.~Washha, and F.~S{\`e}des.
\newblock Online social network phenomena: buzz, rumor and spam.
\newblock In Z.-H. Zhou and C.~Zhang, editors, {\em How information systems can help in alarm/alert detection}, pages 219--239. Elsevier, Oxford, UK, 2018.

\bibitem{muniz2018combining}
C.~P. Muniz, R.~Goldschmidt, and R.~Choren.
\newblock Combining contextual, temporal and topological information for unsupervised link prediction in social networks.
\newblock {\em Knowledge-Based Systems}, 156:129--137, 2018.

\bibitem{peng2022centralized}
S.~Peng, Y.~Yang, M.~Mao, and D.-S. Park.
\newblock Centralized machine learning versus federated averaging: A comparison using mnist dataset.
\newblock {\em KSII Transactions on Internet and Information Systems (TIIS)}, 16(2):742--756, 2022.

\bibitem{rubin2010effect}
B.~Rubin, R.~Fernandes, M.~D. Avgerinou, and J.~Moore.
\newblock The effect of learning management systems on student and faculty outcomes.
\newblock {\em The internet and higher education}, 13(1-2):82--83, 2010.

\bibitem{sahay2023predicting}
R.~Sahay, S.~Nicoll, M.~Zhang, T.-Y. Yang, C.~Joe-Wong, K.~A. Douglas, and C.~G. Brinton.
\newblock Predicting learning interactions in social learning networks: a deep learning enabled approach.
\newblock {\em IEEE/ACM Transactions on Networking}, 31(5):2086--2100, 2023.

\bibitem{saito2015precision}
T.~Saito and M.~Rehmsmeier.
\newblock The precision-recall plot is more informative than the roc plot when evaluating binary classifiers on imbalanced datasets.
\newblock {\em PloS one}, 10(3):1--21, 2015.

\bibitem{salton1975vector}
G.~Salton, A.~Wong, and C.-S. Yang.
\newblock A vector space model for automatic indexing.
\newblock {\em Communications of the ACM}, 18(11):613--620, 1975.

\bibitem{sengupta2024fedel}
D.~Sengupta, S.~S. Khan, S.~Das, and D.~De.
\newblock Fedel: Federated education learning for generating correlations between course outcomes and program outcomes for internet of education things.
\newblock {\em Internet of Things}, 25:101056--101082, 2024.

\bibitem{williams2021social}
D.~Williams-Dobosz, R.~F.~L. Azevedo, A.~Jeng, V.~Thakkar, S.~Bhat, N.~Bosch, and M.~Perry.
\newblock A social network analysis of online engagement for college students traditionally underrepresented in stem.
\newblock In {\em LAK21: 11th International Learning Analytics and Knowledge Conference}, pages 207--215, 2021.

\bibitem{wu2021fedgnn}
C.~Wu, F.~Wu, Y.~Cao, Y.~Huang, and X.~Xie.
\newblock Fedgnn: Federated graph neural network for privacy-preserving recommendation.
\newblock {\em arXiv preprint arXiv:2102.04925}, 2021.

\bibitem{wu2021federated}
J.~Wu, Z.~Huang, Q.~Liu, D.~Lian, H.~Wang, E.~Chen, H.~Ma, and S.~Wang.
\newblock Federated deep knowledge tracing.
\newblock In {\em Proceedings of the 14th ACM international conference on web search and data mining}, pages 662--670, 2021.

\bibitem{wut2021person}
T.-m. Wut and J.~Xu.
\newblock Person-to-person interactions in online classroom settings under the impact of covid-19: a social presence theory perspective.
\newblock {\em Asia Pacific Education Review}, 22(3):371--383, 2021.

\bibitem{xu2024large}
D.~Xu, W.~Chen, W.~Peng, C.~Zhang, T.~Xu, X.~Zhao, X.~Wu, Y.~Zheng, Y.~Wang, and E.~Chen.
\newblock Large language models for generative information extraction: A survey.
\newblock {\em Frontiers of Computer Science}, 18(6):186357--186381, 2024.

\bibitem{xu2018many}
Y.~Xu, C.~F. Lynch, and T.~Barnes.
\newblock How many friends can you make in a week?: Evolving social relationships in moocs over time.
\newblock {\em International Educational Data Mining Society}, 2018.

\bibitem{yang2023palr}
F.~Yang, Z.~Chen, Z.~Jiang, E.~Cho, X.~Huang, and Y.~Lu.
\newblock Palr: Personalization aware llms for recommendation.
\newblock {\em arXiv preprint arXiv:2305.07622}, 2023.

\bibitem{yang2018predicting}
T.-Y. Yang, C.~G. Brinton, and C.~Joe-Wong.
\newblock Predicting learner interactions in social learning networks.
\newblock In {\em IEEE INFOCOM 2018-IEEE Conference on Computer Communications}, pages 1322--1330. IEEE, 2018.

\bibitem{yun2021neo}
S.~Yun, S.~Kim, J.~Lee, J.~Kang, and H.~J. Kim.
\newblock Neo-gnns: Neighborhood overlap-aware graph neural networks for link prediction.
\newblock {\em Advances in Neural Information Processing Systems}, 34:13683--13694, 2021.

\bibitem{zhang2023fedala}
J.~Zhang, Y.~Hua, H.~Wang, T.~Song, Z.~Xue, R.~Ma, and H.~Guan.
\newblock Fedala: Adaptive local aggregation for personalized federated learning.
\newblock In {\em Proceedings of the AAAI Conference on Artificial Intelligence}, volume~37, pages 11237--1fedl2p1244, 2023.

\bibitem{zhao2023survey}
W.~X. Zhao, K.~Zhou, J.~Li, T.~Tang, X.~Wang, Y.~Hou, Y.~Min, B.~Zhang, J.~Zhang, Z.~Dong, et~al.
\newblock A survey of large language models.
\newblock {\em arXiv preprint arXiv:2303.18223}, 1(2), 2023.

\bibitem{zhou2009predicting}
T.~Zhou, L.~L{\"u}, and Y.-C. Zhang.
\newblock Predicting missing links via local information.
\newblock {\em The European Physical Journal B}, 71:623--630, 2009.

\bibitem{zytek2024explingo}
A.~Zytek, S.~Pido, S.~Alnegheimish, L.~Berti-Equille, and K.~Veeramachaneni.
\newblock Explingo: Explaining ai predictions using large language models.
\newblock In {\em 2024 IEEE International Conference on Big Data (BigData)}, pages 1197--1208. IEEE, 2024.

\end{thebibliography}
%\balancecolumns
%\appendix
%Appendix A
% That's all folks!
\end{document}